\documentclass[authoryear,10pt,nopreprintline]{elsarticle}

\usepackage{amsmath}
\usepackage{natbib}
\usepackage{geometry}
\usepackage{graphicx}
\usepackage[hidelinks]{hyperref}
\usepackage{float}
\usepackage{pdfpages}
\DeclareUnicodeCharacter{0301}{\'{e}}
\usepackage{sectsty}
\usepackage{titlesec}
\usepackage{listings}
\usepackage{enumitem}
\usepackage{amssymb}
\usepackage{setspace}
\usepackage{tabularx}
\usepackage{booktabs}
\usepackage{placeins}
\usepackage{float}
\usepackage[utf8]{inputenc}
\usepackage{caption}
\usepackage{subcaption}
\usepackage[title]{appendix}
\newcommand{\dd}{\text{d}}
\usepackage{textcomp}
\usepackage{capt-of}
\usepackage{multirow}

\begin{document}

\begin{frontmatter}

\author[1,2,3]{Elise Mills\corref{cor1}}
\ead{elise.mills@qut.edu.au}
\cortext[cor1]{Corresponding Author}
\author[4,5,6]{Graeme F. Clark}
\ead{graeme.clark@sydney.edu.au}
\author[2]{Matthew J. Simpson}
\ead{matthew.simpson@qut.edu.au}
\author[7]{Mark Baird}
\ead{mark.baird@csiro.au}
\author[1,2,3,8]{Matthew P. Adams}
\ead{mp.adams@qut.edu.au}

\affiliation[1]{organization={Securing Antarctica's Environmental Future},
            addressline={Queensland University of Technology}, 
            city={Brisbane},
            postcode={4001}, 
            state={QLD},
            country={Australia}}
            
\affiliation[2]{organization={School of Mathematical Sciences},
            addressline={Queensland University of Technology}, 
            city={Brisbane},
            postcode={4000}, 
            state={QLD},
            country={Australia}}

\affiliation[3]{organization={Centre for Data Science},
            addressline={Queensland University of Technology}, 
            city={Brisbane},
            postcode={4000}, 
            state={QLD},
            country={Australia}}
            
\affiliation[4]{organization={Evolution \& Ecology Research Centre},
            addressline={University of New South Wales}, 
            city={Sydney},
            postcode={2052}, 
            state={NSW},
            country={Australia}}

\affiliation[5]{organization={Centre of Marine Science and Innovation},
            addressline={University of New South Wales}, 
            city={Sydney},
            postcode={2052}, 
            state={NSW},
            country={Australia}}
            
\affiliation[6]{organization={School of Life and Environmental Sciences},
            addressline={University of Sydney}, 
            city={Camperdown},
            postcode={2006}, 
            state={NSW},
            country={Australia}}

\affiliation[7]{organization={CSIRO Environment},
            city={Hobart},
            postcode={7001}, 
            state={TAS},
            country={Australia}}

\affiliation[8]{organization={School of Chemical Engineering},
            addressline={The University of Queensland}, 
            city={St Lucia},
            postcode={4072}, 
            state={QLD},
            country={Australia}}

\title{A generalised sigmoid population growth model with energy dependence: application to quantify the tipping point for Antarctic shallow seabed algae}

\begin{abstract}
Sigmoid growth models are often used to study population dynamics. The size of a population at equilibrium commonly depends explicitly on the availability of resources, such as an energy or nutrient source, which is not explicit in standard sigmoid growth models. A simple generalised extension of sigmoid growth models is introduced that can explicitly account for this resource-dependence, demonstrated by three examples of this family of models of increasing mathematical complexity. Each model is calibrated and compared to observed data for algae under sea-ice in Antarctic coastal waters. It was found that through careful construction, models satisfying the proposed framework can estimate key properties of a sea-ice break-out controlled tipping point for the algae, which cannot be estimated using standard sigmoid growth models. The proposed broader family of energy-dependent sigmoid growth models likely has usage in many population growth contexts where resources limit population size.
\end{abstract}


\end{frontmatter}

\section{Introduction}\label{sec:intro}

Systems with a dependence on the availability of resources, such as an energy or nutrient source, are ubiquitous in population biology across all flora and fauna: for example, the dependence of plant growth on available light, including minimum light requirements \citep{Erftemeijer2006}, and the dependence of human health on energy intake, including minimum food intake requirements \citep{WHO2004}. With this in mind, we note that many real systems require a steady state population size $N^*>0$ only when an external forcing beneficial to the system, generically here referred to as `energy' $E$, is above a minimum threshold value, known as a `compensation' value $E_{\text{c}}$, and  $N^* = 0$ if these energetic requirements are not met. This is a property that commonly used growth models are unable, without modification, to account for.  

For example, classical sigmoid growth models, such as the logistic growth model and two alternatives to that model, the Gompertz and Richards' models \citep{Tsoularis2002}, have found wide application across multiple disciplines. These models' applications range from biology and ecology, for example modelling populations \citep{Murray1989}, leaf area growth \citep{Thornley1990} and forest recovery \citep{Acevedo2012}, to modelling systems in other fields, such as the market penetration of new products \citep{Fisher1971}. The desired application generally guides the choice of model \citep{Simpson2022}, with the key differences of each model occurring at larger densities \citep{Browning2017}. The basic frameworks of such models and their associated phenomena have been discussed extensively elsewhere \citep{Banks1993}. To summarise, a general sigmoid growth model takes the form:
\begin{equation}
\frac{\dd N}{\dd t} = rN(t) \, f_{\text{c}} \Big( N(t); K, \lambda \Big), \label{eq:sigmoid_generic}
\end{equation}
where $r > 0$ is the growth rate, $N(t) \geq 0$ is the population size, $t \geq 0$ is time  and  $f_{\text{c}}(N(t); K, \lambda) \geq 0$ is a non-dimensional `crowding' function, with carrying capacity $K$ and other parameters $\lambda$ (where relevant), that influences the population's net growth rate \citep{Jin2016}, where typically $f_{\text{c}} \Big( K; K, \lambda \Big) = 0$. These crowding functions take on the following typical forms, which are governed by the specific model being used:
\begin{subequations}
\begin{align}
f_{\text{c}}(N(t); K) &= 1 - \frac{N(t)}{K} & \text{(Logistic growth)} \label{eq:crowd_logistic}\\
f_{\text{c}}(N(t); K, \beta) &= 1 - \left( \frac{N(t)}{K} \right)^\beta & \text{(Richards' growth)}\label{eq:crowd_richards} \\
f_{\text{c}}(N(t); K) &= \log \left( \frac{K}{N(t)} \right) & \text{(Gompertz growth)}\label{eq:crowd_gompertz}
\end{align}
\label{eq:crowding}%
\end{subequations}
where $K>0$ (and $\beta > 0$ when present, as the only member of additional parameters $\lambda$). The crowding functions listed in Equation \eqref{eq:crowding} result in the stable steady state $N^* = K$ and the unstable steady state $N^* = 0$. Sigmoid growth models with logistic and Richards' crowding functions are characterised by approximately exponential growth $\dd N(t) / \dd t \approx r N(t)$ for small populations ($N(t) \ll K$), and for all crowding functions the growth decreasing to zero as the population size approaches the carrying capacity ($N(t) \to K^-$) due to increasing resource limitation \citep{Murray1989}. The stability of $N^*=K$ means that as long as the initial population size is greater than zero, $N(t) \to K^-$ as $t \to \infty$ \citep{Murray1989}. 

Richer and more nuanced population behaviours arise through strategic modifications of Equation \eqref{eq:sigmoid_generic}. For example, the effect of the crowding function reducing growth as $N(t)>0$ increases towards $K$ \citep{Jin2016} can be partially countered by modifying Equation \eqref{eq:sigmoid_generic} with a weak Allee effect \citep{Allee1932, Wang2002, Taylor2005, Fadai2020}:
\begin{equation}
\frac{\dd N}{\dd t} = rN(t) f_{\text{c}}(N(t); K, \lambda) \left(1 + \frac{N(t)}{C} \right), \label{eq:Allee_weak}
\end{equation}
where $C>0$ is a positive constant. This modification changes the magnitude of $\dd N(t) / \dd t$, but does not alter its steady state behaviour. 

Alternatively, Equation \eqref{eq:sigmoid_generic} can be modified so that $N^* = 0$ and $N^* = K$ are \textit{both} stable steady states, via introduction of a strong Allee effect \citep{Allee1932, Wang2002, Taylor2005, Fadai2020}:
\begin{equation}
\frac{\dd N}{\dd t} = rN(t) f_{\text{c}}(N(t); K, \lambda) \left(\frac{N(t)}{D} - 1 \right), \label{eq:Allee_strong}
\end{equation}
where $D>0$ is a positive constant. The concept of a second stable steady state is an example of bistability \citep{Holling1973, NoyMeir1975}. 

The models discussed thus far only permit $N^* = 0$ and/or $N^* = K$ as stable steady states. However, in practice, populations may stabilise at population sizes or abundances that depend on external forcing; for example, $N^*$ may gradually increase with environmental suitability, and/or $N^*$ may gradually decrease with environmental stress. Additional modifications to Equation \eqref{eq:sigmoid_generic} are required to simulate this behaviour.

One approach to obtain $N^* \neq 0,K$ is the introduction of limiting or harvesting terms, which describe the removal of portions of a population \citep{Brauer1975, Brauer1979}, for example, the effects of fishing on a fish population \citep{Murray1989}. Such a term can be constant or variable depending on, for example, time \citep{Idlango2017} or density \citep{Cooke1986}, and is incorporated as shown in Equation \eqref{eq:harvesting_gen} for a time-dependent harvesting term $H(t)$:
\begin{equation}
	\frac{\dd N(t)}{\dd t} = r N(t) \, f_{\text{c}}\left( N(t); K, \lambda \right) - H(t), \label{eq:harvesting_gen}
\end{equation}
where it is typically assumed that $H(t) \geq 0$. For example, consider the application of the harvesting term to the logistic model:
\begin{equation}
	\frac{\dd N(t)}{\dd t} = r N(t) \Big( 1 - \frac{N(t)}{K} \Big) - H(t). \label{eq:harvesting_log}
\end{equation}
Setting $H(t) = H_0 N(t)$, for some constant $H_0 > 0$, changes the model's dynamics so that the stable steady state is $N^* = K(1 - H_0 / r)$ if $H_0 < r$, and the stable steady state is $N^* = 0$ if $H_0 > r$. Thus, the stable steady state of the population $N^*$ decreases as the per-capita harvesting constant $H_0$ increases. Different steady state behaviour can arise from different choices of $H(t)$, although this behaviour may not be obvious or apparent directly from Equation \eqref{eq:harvesting_log} without additional derivations.

To the same end, other works \citep{Grozdanovski2009, Idlango2012,  Shepherd2012, Idlango2017} consider sigmoid growth models with $r$ and $K$ as functions of time, $r(t)$ and $K(t)$:
\begin{equation}
\frac{\dd N(t)}{\dd t} = r(t) N(t) \, f_{\text{c}} \Big( N(t); K(t), \lambda  \Big). \label{eq:functime}
\end{equation}
Such modifications allow the incorporation of more dynamic effects, such as responses of the population to background seasonality \citep{Grozdanovski2009}. Although these modifications do take into account environmental factors, the only allowed steady states are $N^*=0$ or the quasi-steady state $N=K(t)$ (or within a small neighbourhood of $N=K(t)$, see \citet{Idlango2012, Shepherd2012, Idlango2017}) with no possibility of $0 < N^* \ll K$. 

One potential application of Equation \eqref{eq:functime} is to treat the growth rate $r(t)$ and/or the carrying capacity $K(t)$ as functions of an external forcing that is permitted to be time-varying. If this external forcing is detrimental to the population, the resulting effects may be similar to that of the harvesting rate in Equation \eqref{eq:harvesting_gen}; we have not yet dealt with the case where the external forcing is beneficial to the population. In the case of a beneficial external forcing, we may require, for example, that $N^*>0$ only when a relevantly defined quantity of energy $E$ exceeds some compensation value $E_{\text{c}}$, and $N^* = 0$ if this energy requirement is not met (i.e.\ if $E<E_{\text{c}}$). In ecological situations, a limiting resource or energy $E$ can be time-dependent ($E(t)$), but we will limit our considerations here to examples where energy $E$ is constant over time.

A naive modification of Equation \eqref{eq:functime} to capture energy dependence satisfying the above requirements ($N^*>0$ when $E>E_{\text{c}}$, $N^*=0$ when $E<E_{\text{c}}$) is to choose $K(E)$ and $r(E)$ to both be monotonically non-decreasing (i.e.\ either static or increasing) functions of $E$, such that $K(E_{\text{c}}) = 0$, $r(E_{\text{c}}) = 0$, and:
\begin{equation}
\lim_{E \to E_{\text{c}}} \Big[ r \left(E\right) \, f_{\text{c}} \left( N(t); K(E), \lambda \right) \Big] =0, \label{eq:lim_Ec0}
\end{equation}
with this latter condition required to avoid singularity in $\dd N / \dd t$ when $E = E_{\text{c}}$. For example, for logistic growth, Equation \eqref{eq:lim_Ec0} requires $\lim_{E \to E_{\text{c}}}r(E)$ to approach zero faster than $\lim_{E \to E_{\text{c}}}K(E)$ to avoid singularity. Equation \eqref{eq:lim_Ec0} is thus a mathematical inconvenience that could also yield non-realistic behaviour for $\dd N / \dd t$ in the vicinity of $E = E_{\text{c}}$ if $r(E)$ and $K(E)$ are not chosen carefully. 

An alternative inelegant approach to modelling energy dependence is to use Equation \eqref{eq:harvesting_log}, noting that $H(t) = H_0 N(t)$ permits a smooth transition from $N^* = 0$ when $H_0 > r$ to $N^* = K(1 - H_0 /r)$ when $H_0 < r$, but only if $H_0$ is interpreted as an environmental stress (i.e.\ conceptually the opposite of the energy dependence we wish to capture). If we wish for the model to capture steady state population behaviour that has energy dependence, the structure of Equation \eqref{eq:harvesting_log} does not intuitively capture this concept. 

One application where growth models that capture energy dependence may be of great use is benthic ecosystems in the coastal waters of Antarctica. Antarctic benthic ecosystems play an important role in many global cycles \citep{Murphy2021}; however there are many challenges faced when attempting to model these ecosystems \citep{Kennicutt2014}. Critically, detailed long-term data is often not available \citep{Convey2014, Koerich2023}, driving the need for modelling that incorporates ecological processes \citep{Koerich2023} or environmental conditions, such as light availability. 

Coastal waters surrounding Antarctica are subject to highly seasonal variation in sea-ice cover, the presence of which restricts the transmission of light to the benthic environment below \citep{Katlein2015,Clark2017}. The total time period of sea-ice cover per year in Antarctic waters is generally decreasing \citep{Abram2010, Stammerjohn2012}, and models predict decreases in the extent of sea-ice into the future \citep{Mayewski2009}. These processes result in an increase in the amount of light per year reaching the benthic ecosystems, an effect exacerbated by the non-linear relationship between the timing of sea-ice break-out and annual light dose \citep{Runcie2006, Clark2013}. More specifically, the light dose per day depends on what day of the year it is -- as both the day length and height of the Sun in the sky determines the daily light dose; this yields non-linearity in the relationship between the timing of sea-ice break-out and annual light dose. 

The total time period of sea-ice cover each year has a heavy influence on Antarctic shallow-water marine ecosystems, whose composition is determined by the amount of light received \citep{Miller2008, Clark2017}. These ecosystems are home to a diverse number of species \citep{Griffiths2010}, with a high level of endemism (i.e.\ species uniqueness) \citep{Dell1972, Clayton1994, Peck2018}. The presence of sea-ice creates a dark and calm environment which favours habitation by macroinvertebrate communities \citep{Picken1985, Clayton1994, Clark2013, Clark2015}. However, should the period of sea-ice cover decrease to the extent that the ecosystem is receiving sufficient light to allow algae growth, the algae will outcompete invertebrates, resulting in a tipping point where endemic invertebrate communities are replaced with beds of macroalgae \citep{Clark2013}. Such trends have already been observed in the Arctic \citep{Kortsch2012, KrauseJensen2012, Scherrer2019} and Antarctic \citep{Quartino2013}, and pose a significant risk to the survival of invertebrate communities and thus the biodiversity of the Antarctic continent's shallow-water benthic zone. To the best of our knowledge, these ecosystems are the only Antarctic ecosystems assessed by the International Union for Conservation of Nature Red List of Ecosystems (IUCN RLE) for their collapse threat status  \citep{Clark2015,IUCNCEM}, having been classified as ranging from Near Threatened to Vulnerable, and are at further risk due to record low sea-ice coverage in 2023 \citep{Purich2023}.

In this work, we introduce a generalised sigmoid population growth model framework that intuitively captures energy dependence of the steady state population. The family of models we introduce is a straightforward extension of sigmoid population growth models that can capture a wide range of possible functional forms. These models guarantee a steady state population of $N^*=0$ when the available energy $E$ is less than a threshold value ($E_{\text{c}}$), and a steady state population that is a non-decreasing function of $E$ when $E$ is above the threshold value $E_{\text{c}}$. We present and analyse three examples of this model family, of increasing complexity and relevance. 

We apply and calibrate the three example models to a case study involving data collected for Antarctic shallow water benthic communities \citep{Clark2013} at one time point per community; this data describes the balance between algae and invertebrate populations (discussed in detail in Section \ref{sec:data_sources}). We compare the merits of our proposed energy-dependent models for forecasting algae cover in response to available light, and for predicting tipping points and other key ecological parameters in benthic algae-invertebrate ecosystems where light availability is restricted by sea-ice cover.

The case study is an example of a system in which changes in the abundance of these energy resources beyond a critical threshold, or `tipping point', can result in sudden and widespread change of the system to an alternative state \citep{Scheffer2001}. However, more broadly, the presented models are sufficiently general that they can represent the dynamics of any system in which there is nontrivial dependence of a steady state population on available energy resources (for example nutrients, light, prey species, habitat).

\section{Methods}

\subsection{Theory: Energy-dependent sigmoid growth models}\label{sec:growth_models}

The generalised sigmoid population growth model with energy dependence that we introduce is:
\begin{equation}
\frac{\dd N(t)}{\dd t} = r N(t)\, f_{\text{c}}(N(t); K, \lambda)\, \Big( f_{\text{E}} (E; \lambda) - f_{\text{N}} (N(t); \lambda)\, \Big), \label{eq:log_mod}
\end{equation}
where $N(t)$ is the population size satisfying $0 \leq N(t) \leq K$, $r > 0$ is a nominal growth rate, $E \geq 0$ is the energy available to the population, $f_{\text{c}}(N(t); K, \lambda) \geq 0$ is a `crowding' function, $f_{\text{E}}(E; \lambda)$ is an `energy dependence' function, $f_{\text{N}}(N(t);\lambda) \geq 0$ is a `population penalty' function, $K > 0$ is a population carrying capacity and $\lambda$ represents any other constant parameters (where relevant). 

\begin{figure}[ht]
\captionsetup{skip=0\baselineskip}
\captionsetup[subfigure]{skip=0pt, singlelinecheck=false}
	\centering
	\begin{subfigure}[b]{0.45\textwidth}
		\centering
		\caption{Functions $f_{\text{N}}$ and $f_{\text{c}}$}\includegraphics[width = \textwidth]{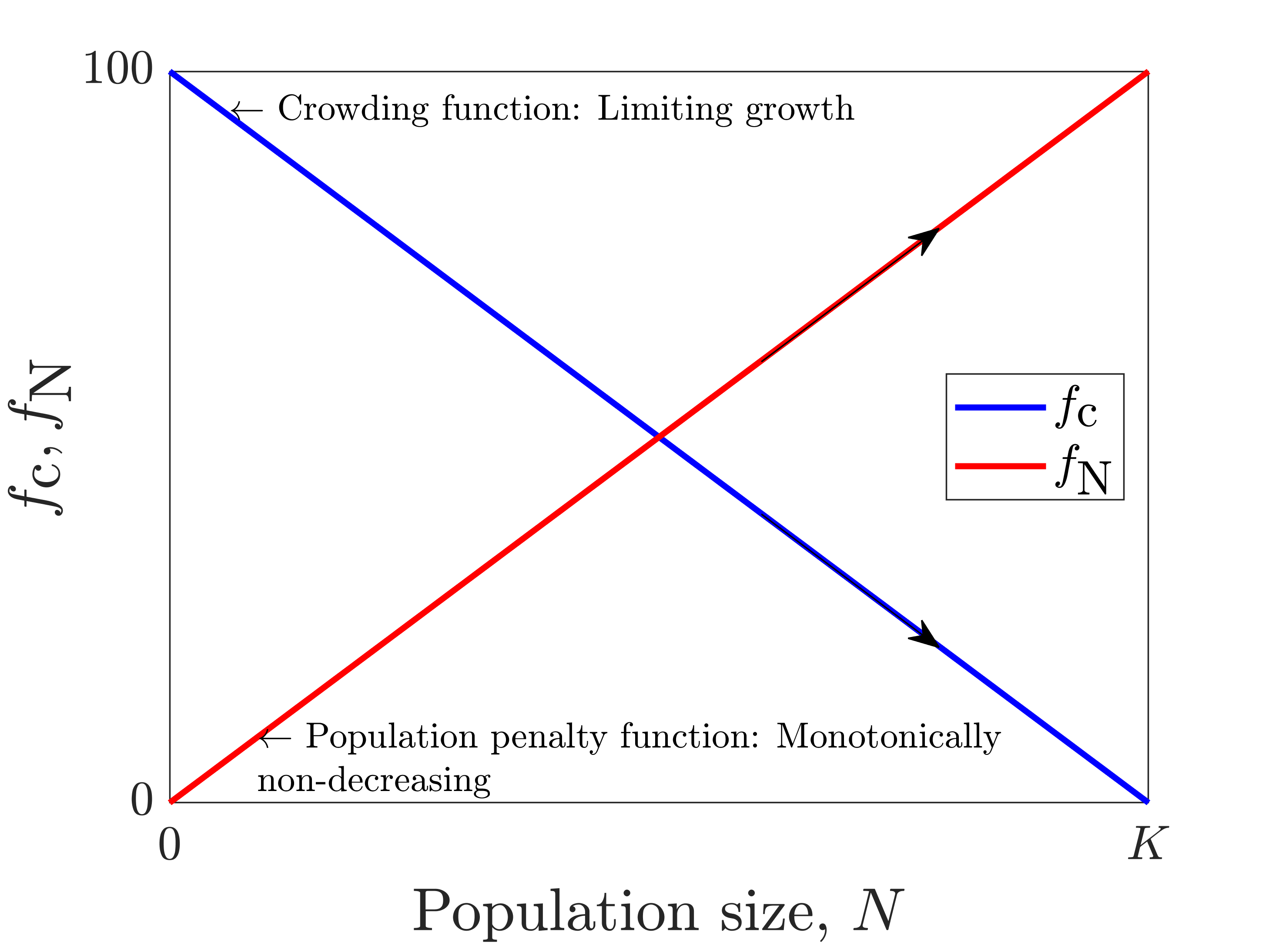}
		\label{fig:fnfc}
	\end{subfigure}
	\begin{subfigure}[b]{0.45\textwidth}
		\centering
		\caption{Function $f_{\text{E}}$}\includegraphics[width = \textwidth]{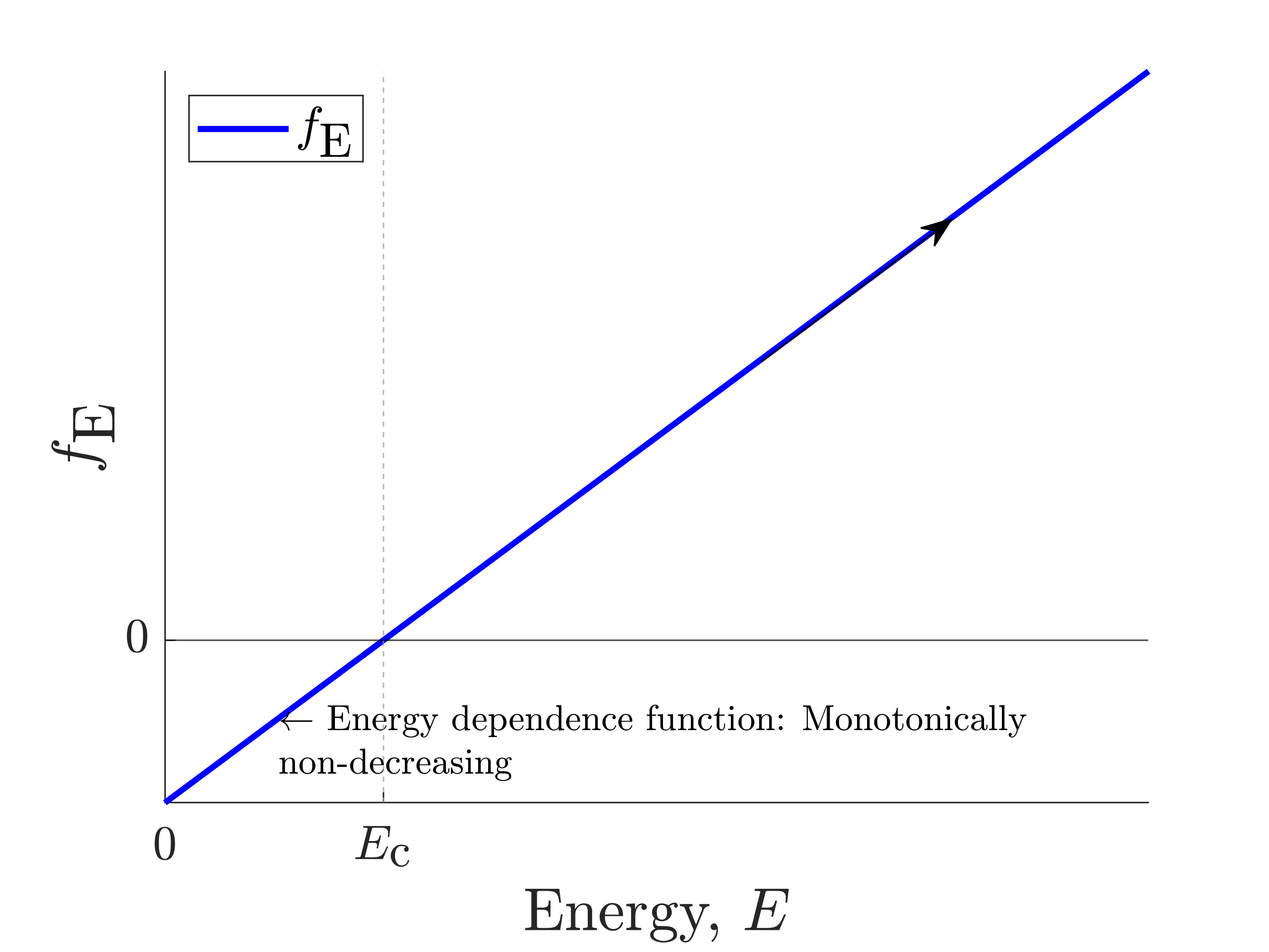}
		\label{fig:fe}
	\end{subfigure}
	\caption{Visual depiction of one potential choice for the functions $f_{\text{N}}$, $f_{\text{c}}$ and $f_{\text{E}}$ to be used in the proposed energy-dependent sigmoid growth model, Equation \eqref{eq:log_mod}. The specific functions $f_{\text{N}}$, $f_{\text{c}}$ and $f_{\text{E}}$ depicted here are the linear relationships that form the second model example we introduce, discussed further in Section \ref{sec:model_2}.}
	\label{fig:functions}
\end{figure}

The crowding function $f_{\text{c}}(N(t); K, \lambda) \geq 0$ limits the population's net growth rate as $N(t)$ approaches the population carrying capacity $K$ \citep{Jin2016, Simpson2022}. The energy dependence function $f_{\text{E}}(E; \lambda)$ can take positive or negative values and is a monotonically non-decreasing function of $E$ such that it is zero when the energy is equal to some `compensation' value $E_{\text{c}}$, i.e.\ $f_{\text{E}}(E_{\text{c}}; \lambda) = 0$. The population penalty function $f_{\text{N}}(N(t);\lambda) \geq 0$ is intended to represent reduced energy efficiency as the population $N(t)$ increases. To represent a penalty rather than a gain, $f_{\text{N}}(N(t); \lambda)$ must be a monotonically non-decreasing function of $N(t)$. Additionally, it follows that $f_{\text{N}}(N(t); \lambda)$ is zero when the population is zero, $f_{\text{N}}(0;\lambda) = 0$. Graphical representations of one potential choice for the functions $f_{\text{N}}$, $f_{\text{c}}$ and $f_{\text{E}}$ are shown in Figure \ref{fig:functions}; while linear examples are shown, these functions can also be non-linear.  Equation \eqref{eq:log_mod} is an extension of the general sigmoid growth model (Equation \eqref{eq:sigmoid_generic}) via inclusion of the additional factor $(f_{\text{E}} (E; \lambda) - f_{\text{N}} (N(t); \lambda))$ to incorporate energy dependence; this factor is analogous to an Allee effect (Section \ref{sec:intro}). Note that Equation \eqref{eq:log_mod} is only defined for a population $N$ satisfying $0 \leq N(t) \leq K$ because a population cannot be negative, and we wish to avoid non-physical behaviour when $N(t) > K$ in the case where $f_{\text{c}}(N(t); K, \lambda) < 0$ and $(f_{\text{E}}(E; \lambda) - f_{\text{N}}(N(t); \lambda)) < 0$ as such a situation could yield unrestricted growth. 

\FloatBarrier

In the following sections, we explain and examine three models adhering to the form of Equation \eqref{eq:log_mod} to demonstrate the generality of the introduced model family. In these models we will define $f_{\text{E}}(E ; \lambda)$ and $f_{\text{N}}(N(t) ;  \lambda)$, but we will keep the crowding function $f_{\text{c}}(N(t) ; K, \lambda)$ general where possible, since forms of $f_{\text{c}}$ have been readily discussed elsewhere (e.g.\ \citealt{Simpson2022}).

Possible steady state values of Equation \eqref{eq:log_mod} are $N^*=0$, $N^*=K$ or $N^*$ satisfying $f_{\text{E}}(E; \lambda) = f_{\text{N}}(N(t); \lambda)$, for any appropriately chosen crowding function $f_{\text{c}}(N(t); K, \lambda)$, such as those listed in Equation \eqref{eq:crowding}. Because $f_{\text{E}}(E; \lambda)$ and $f_{\text{N}}(N(t); \lambda)$ are both monotonically non-decreasing functions, this implies the inverse function $f_{\text{N}}^{-1}(f_{\text{E}}(E; \lambda))$ is also a monotonically non-decreasing function, so the steady state $N^*$ satisfying $f_{\text{E}}(E; \lambda) = f_{\text{N}}(N(t); \lambda)$ will be a monotonically non-decreasing function of $E$. Also note, for logistic and Richards' growth models, $r$ could be interpreted as an approximation of the per-capita growth rate in the specific case of when both $N(t)$ is small and $f_{\text{E}}(E; \lambda) \approx 1$. 

\subsubsection{Example model 1: Linear energy dependence and no population penalty}\label{sec:model_1}

\begin{figure}[h!]
	\centering
		\includegraphics[width = 0.6\textwidth]{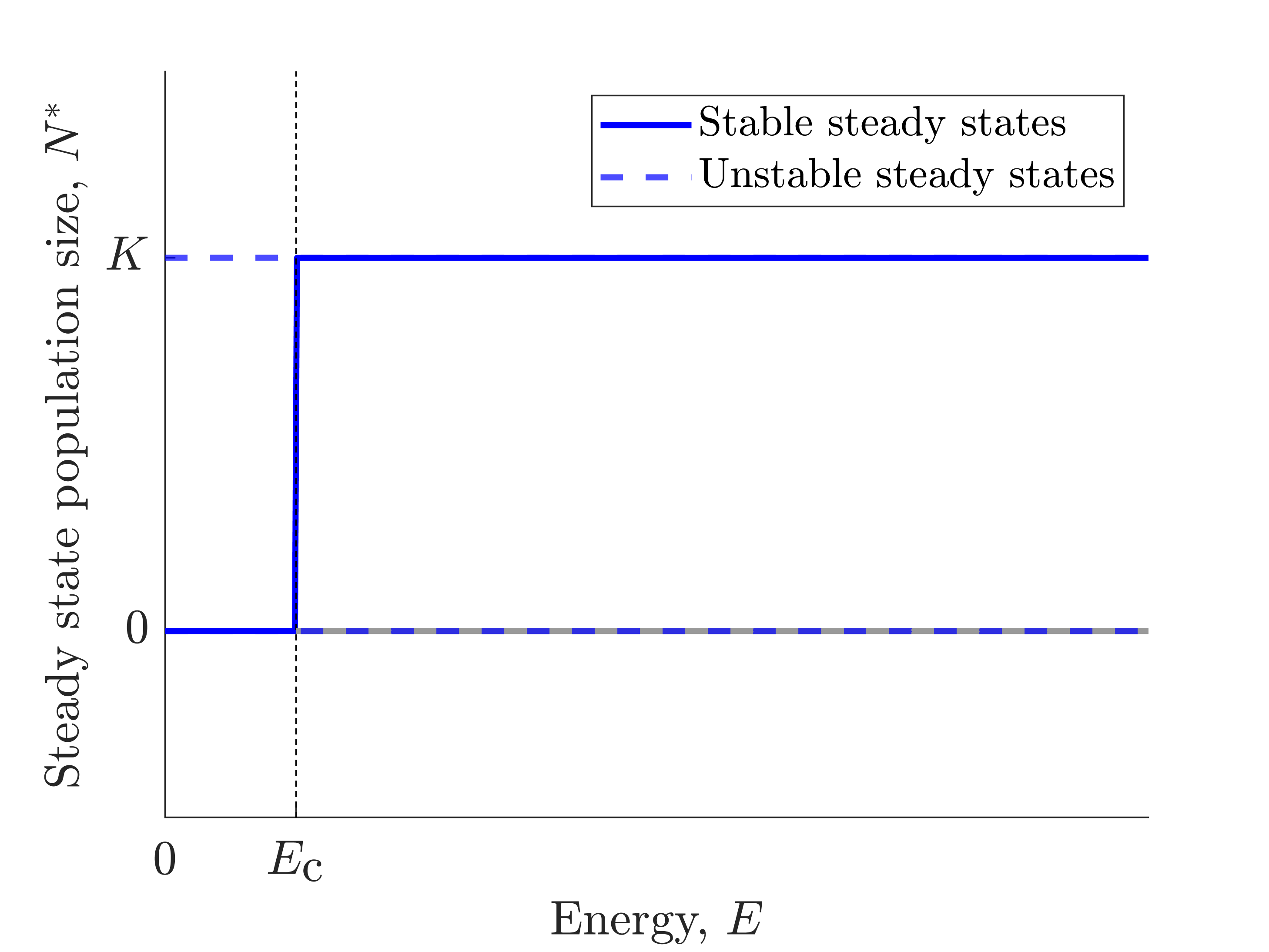}
		\caption{Steady states of the energy-dependent sigmoid growth model with linear energy dependence and no population penalty (Step Model, defined in Equation \eqref{eq:model_1}).}
		\label{fig:model_1_ss}
\end{figure}

The first model adhering to the form of Equation \eqref{eq:log_mod} we discuss is hereafter referred to as the `Step Model', described as such due to its steady state behaviour shown in Figure \ref{fig:model_1_ss}. The Step Model has linear energy dependence, where $f_{\text{E}} (E; \lambda) = (E - E_{\text{c}})/E_{\text{c}}$, and no population penalty $f_{\text{N}} (N(t); \lambda) = 0$:
\begin{equation}
\frac{\dd N(t)}{\dd t} = r N(t) f_{\text{c}}(N(t); K, \lambda) \left( \frac{E - E_{\text{c}}}{E_{\text{c}}} \right), \label{eq:model_1}
\end{equation}
as this yields perhaps the simplest nontrivial form of Equation \eqref{eq:log_mod} whereby $f_{\text{E}}(E;\lambda)$ and $f_{\text{N}}(N(t); \lambda)$ satisfy the requirements of (1) both functions being monotonically non-decreasing, (2) the energy dependence function being zero at the compensation energy, $f_{\text{E}} (E; \lambda) = 0$, and (3) the population penalty function being zero when there is zero population $f_{\text{N}} (N(t); \lambda) = 0$. Regardless of whether the crowding function represents logistic, Richards' or Gompertz growth, Equation \eqref{eq:model_1} has stable steady states of $N^*=0$ when $E < E_{\text{c}}$, and $N^*=K$ when $E > E_{\text{c}}$. This model therefore only allows the trivial steady states $N^* = 0$ and $N^* = K$, depicted in Figure \ref{fig:model_1_ss}. The model's behaviour when $E = E_{\text{c}}$ is of little interest, since in practical biology situations, this precise relationship will be highly unlikely to occur. 

\FloatBarrier

The model defined in Equation \eqref{eq:model_1} and its behaviour is analogous to combining the growth rate $r$ and energy dependence $f_{\text{E}}(E; \lambda)$ into an energy-dependent growth rate $r(E) \leftarrow r \times (E-E_{\text{c}})/E$, yielding similarity to time-dependent growth rates \citep{Grozdanovski2009, Idlango2012, Shepherd2012, Idlango2017} and resource- or stressor-dependent growth rates \citep{Baird2016, Turschwell2022}. However, it is interesting to note that, despite Equation \eqref{eq:model_1} having a linear energy dependence for growth, this does not translate into a linear dependence of the steady states $N^*$ on energy $E$ (Figure \ref{fig:model_1_ss}).

\subsubsection{Example model 2: Linear energy dependence and linear population penalty}\label{sec:model_2}

\begin{figure}[h!]
	\centering
		\includegraphics[width = 0.6\textwidth]{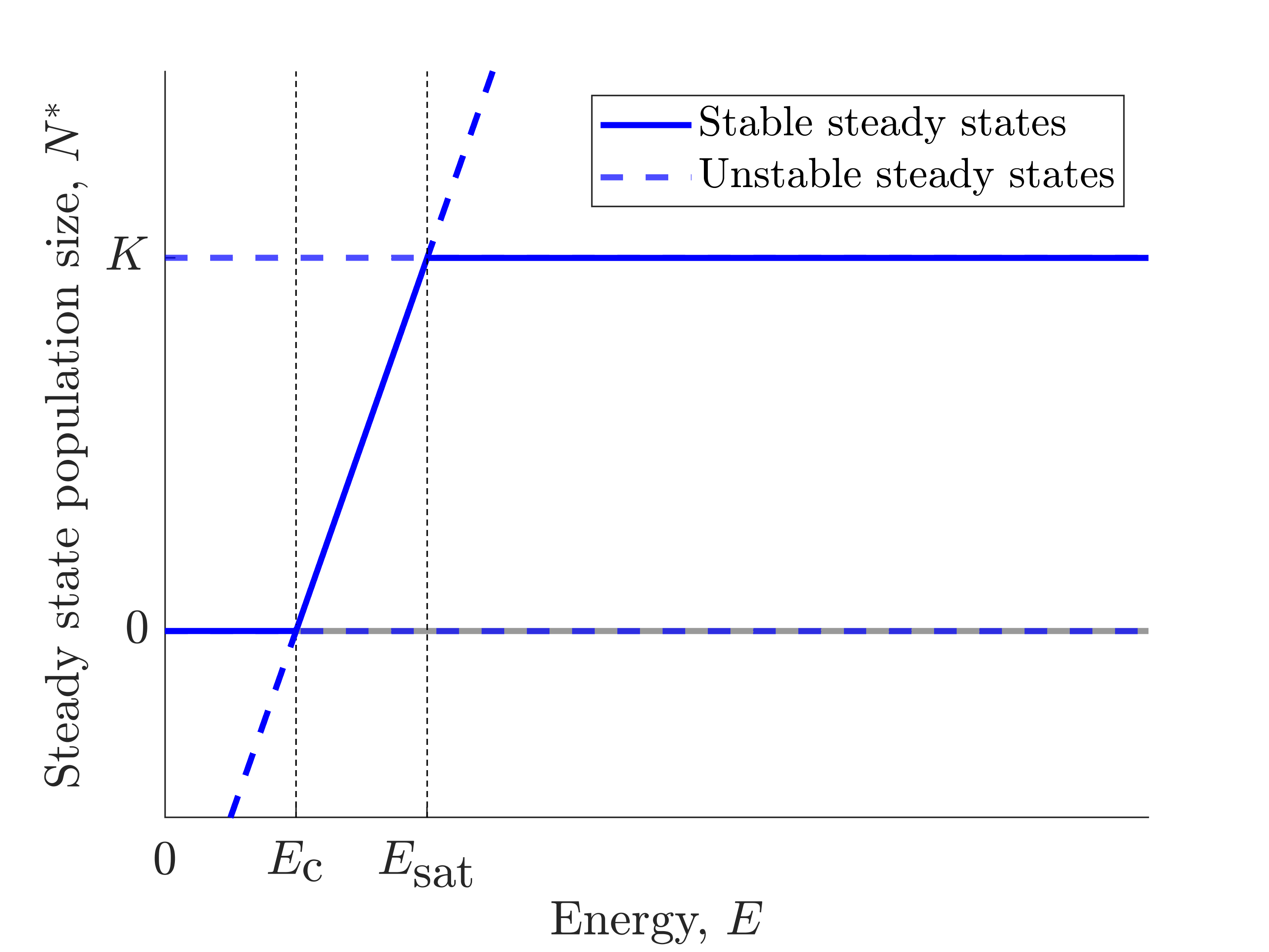}
	\caption{Steady states of the energy-dependent sigmoid growth model with linear energy dependence and linear population penalty (Linear Model, defined in Equation \eqref{eq:model_2}).}
	\label{fig:model2_ss_analysis_dim}
\end{figure}

The steady state population can become a continuous function of energy if the trivial population penalty function $f_{\text{N}} ( N(t); \lambda )$ chosen in Equation \eqref{eq:model_1} is changed to a slightly less trivial population penalty function $f_{\text{N}} ( N(t); \lambda ) = \kappa N(t)$:
\begin{equation}
\frac{\dd N(t)}{\dd t} = r N(t) f_{\text{c}}(N(t); K, \lambda) \left( \frac{E - E_{\text{c}}}{E_{\text{c}}} - \kappa N(t) \right), \label{eq:model_2}
\end{equation}
where $\kappa > 0$ so that $f_{\text{N}} ( N(t); \lambda )$ is monotonically non-decreasing. This model is hereafter referred to as the `Linear Model', described as such due to its steady state behaviour shown in Figure \ref{fig:model2_ss_analysis_dim}.

Setting $\kappa = (E_{\text{sat}} - E_{\text{c}})/(E_{\text{c}} K)$ for convenience, where $E_{\text{sat}} > E_{\text{c}}$ is a `saturation' value of energy, we find that there are stable steady states  $N^*=0$ when $E \leq E_{\text{c}}$, $N^* = K(E-E_{\text{c}})/(E_{\text{sat}}-E_{\text{c}})$ when $E_{\text{c}} \leq E \leq E_{\text{sat}}$, and $N^* = K$ when $E \geq E_{\text{sat}}$ (Figure \ref{fig:model2_ss_analysis_dim}). We caution that this saturation value $E_{\text{sat}}$ differs from typical ecological interpretations of saturation energy; in the present context, $E_{\text{sat}}$ refers to the value of energy at which the population reaches the maximum possible value above which no further growth is possible. Thus, the steady state population $N^*$ depends linearly on energy $E$ when this energy both exceeds a `compensation' value $E_{\text{c}}$ and is less than a maximum `saturation' value $E_{\text{sat}}$. Equation \eqref{eq:model_2} therefore provides a simple phenomenological model, with easily interpretable parameters, for representing linear energy dependence of steady state populations.
\FloatBarrier

\subsubsection{Example model 3: Logistic growth, linear energy dependence, and nonlinear population penalty derived from geometric arguments for plant cover versus light availability}\label{sec:model_3}

For the third model examined here, we take a rather different approach to the previous two models; this model is hereafter referred to as the `Geometric Model' due to its derivation discussed later in this section. Both the Step Model and Linear Model (Equations \eqref{eq:model_1} and \eqref{eq:model_2} respectively) are phenomenological in origin, i.e.\ they are designed intentionally to obtain the desired model characteristics. However, there are multiple ways in which more complicated models could be constructed that adhere to the form of Equation \eqref{eq:log_mod}. Multiple authors advocate for models that are less phenomenological than the two models described in the previous two sections, and rather desire models based on mechanistic understanding \citep{White2019}. Hence, as a demonstration of the potentially wide application of Equation \eqref{eq:log_mod}, we here show (surprisingly) that a mechanistic ecological model predicting plant cover responses to light availability derived from geometric arguments \citep{Baird2016} adheres to the form of Equation \eqref{eq:log_mod}, when the model's population dynamics are defined in units of ground area covered by the plant. The application of geometric approaches to develop models for plant species processes and growth is a well-established practice \citep{Eagleson2002} but is less common for population dynamics of other species. 

The geometric derivation of this model is advantageous because it expresses dynamics based on a mechanistic understanding; yet it requires only two parameters $r$ and $E_{\text{c}}$ whilst also still yielding dependence of steady state populations on energy. Thus, this model retains the benefits of both the Step Model and Linear Model, but also has a mechanistic basis. This model is already being used to represent the dynamics of multiple marine benthic species in a sophisticated environmental modelling suite that is currently being applied to the Great Barrier Reef \citep{Baird2020}.

\paragraph{Derivation of the Geometric Model}\mbox{}\smallskip

The primary geometric assumption of the Geometric Model is that `effective' ground cover occupied by the population, $A_{\text{eff}}(t)$, which is bounded between zero (0\% cover) and one (100\% cover), increases with population biomass $B(t)$ proportionally to the unoccupied ground area $1 - A_{\text{eff}}(t)$ \citep{Baird2016}:
\begin{equation}
\frac{\dd A_{\text{eff}}(t)}{\dd B(t)} \propto 1 - A_{\text{eff}}(t). \label{eq:baird_dadb}
\end{equation}
Equation \eqref{eq:baird_dadb} represents the growth of a plant population, where new leaves are randomly placed continuously on the ground area; a fraction $1 - A_{\text{eff}}(t)$ of these new leaves will cover unoccupied ground area \citep{Baird2020}. Following \cite{Baird2016}, we next define the leaf area $A(t)$ to represent the total area of all leaves, which is a different value to the effective area as viewed from above, $A_{\text{eff}}(t)$. Because, when viewed from above, leaves may cover other leaves, $A(t) \geq A_{\text{eff}}(t)$, with equality only if no leaf is covering any other leaf. Then, mathematically assuming $A_{\text{eff}}(t)=0$ implies $B(t)=0$, and introducing $\Omega$ as the ratio of plant leaf area to biomass, $\Omega = A(t)/B(t)$, one can obtain from Equation \eqref{eq:baird_dadb} that:
\begin{equation}
A_{\text{eff}}(t) = 1 - e^{-\Omega B(t)}, \label{eq:baird_Aeff}
\end{equation}
as shown in Appendix B of \cite{Baird2016}. 

Since Equation \eqref{eq:baird_Aeff} considers a plant population, the primary limitation to energy intake (and subsequent growth) is availability of light for absorption by plant leaves. However, not all plant leaves can absorb light because they may be shaded by other leaves. Thus, for this model it is assumed that the increase $\dd B(t)/\dd t$ in population biomass $B(t)$ is proportional to the energy absorbed per ground area occupied by the population, $E A_{\text{eff}}(t)$, minus the minimum (`compensation') energy needed to fuel the total population biomass, $E_{\text{c}} A(t) \equiv E_{\text{c}} \Omega B(t)$:
\begin{equation}
\frac{\dd B(t)}{\dd t} \propto \left[ EA_{\text{eff}}(t) - E_{\text{c}} \Omega B(t) \right]. \label{eq:dbdt_inc}
\end{equation}
Writing Equation \eqref{eq:dbdt_inc} solely in terms of $B(t)$, i.e.\ without dependence on $A_{\text{eff}}(t)$, by substituting Equation \eqref{eq:baird_Aeff}, yields:
\begin{equation}
\frac{\dd B(t)}{\dd t } \propto \left[ E \left( 1-e^{-\Omega B(t)} \right) - E_{\text{c}}\Omega B(t) \right] \label{eq:biomass_ROC}.
\end{equation}
Equation \eqref{eq:biomass_ROC} is the foundation of sophisticated models of seagrass, corals and macroalgae  used in \cite{Baird2016, Baird2020}, although see Appendix B.1 of \cite{Adams2018} for a partial deconstruction of these complicated models towards a form analogous to Equation \eqref{eq:biomass_ROC}. These previous works considered plant population in units of biomass \citep{Baird2016, Baird2020} or leaf area index \citep{Adams2018}. In the present work, we rewrite Equation \eqref{eq:biomass_ROC} in units of plant-occupied ground cover via application of the chain rule:
\begin{align}
\frac{\dd A_{\text{eff}}(t)}{\dd t} &= \frac{\dd A_{\text{eff}}(t)}{\dd B(t)} \frac{\dd B(t) }{\dd t}. \label{eq:Baird_daeff_dt}
\end{align}

Setting $r / E_{\text{c}}$ as the constant of proportionality in Equation \eqref{eq:biomass_ROC}, and combining the resulting equation with Equations \eqref{eq:baird_Aeff} and \eqref{eq:Baird_daeff_dt}, yields:
\begin{align}
\frac{\dd A_{\text{eff}}(t)}{\dd t} &= r A_{\text{eff}}(t) (1-A_{\text{eff}}(t)) \left[ \frac{E-E_{\text{c}}}{E_{\text{c}}} - \frac{\ln\left((1-A_{\text{eff}}(t))^{-1}\right) - A_{\text{eff}}(t)}{A_{\text{eff}}(t)} \right]. \label{eq:Aeff_ROC}
\end{align}

Comparing Equations \eqref{eq:log_mod} and \eqref{eq:Aeff_ROC}, we see that the derived Geometric Model (Equation \eqref{eq:Aeff_ROC}) is a particular case of the energy-dependent sigmoid growth model introduced in Equation \eqref{eq:log_mod}, where:
\begin{align}
N(t) &= A_{\text{eff}}\\
K &= 1 \label{eq:K}\\
f_{\text{c}}(N(t); K, \lambda) &= 1 - \frac{N(t)}{K}\\
f_{\text{E}}(E; \lambda) &= \frac{E - E_{\text{c}}}{E_{\text{c}}} \label{eq:f2E}\\
f_{\text{N}} (N(t); \lambda) &= \frac{K \ln\left((1-N(t)/K)^{-1}\right) - N(t)}{N(t)}. \label{eq:f3A}
\end{align}
Thus, we rewrite here the Geometric Model, for consistency with the other models, in the more general form:
\begin{equation}
\frac{\dd N(t)}{\dd t} = r N(t) \left( 1 - \frac{N(t)}{K} \right) \left[ \frac{E - E_{\text{c}}}{E_{\text{c}}} - \frac{K\ln\left((1-N(t)/K)^{-1}\right) - N(t)}{N(t)} \right]. \label{eq:model_3}
\end{equation}
The crowding function $f_{\text{c}}(N(t); K,\lambda)$ of the Geometric Model is logistic growth (Equation \eqref{eq:crowd_logistic}), and the energy dependence function $f_{\text{E}}(E;\lambda)$ of the Geometric Model is the same as the Step Model and Linear Model, which already satisfies the required properties of being a monotonically non-decreasing function of $E$ and $f_{\text{E}}(E_{\text{c}}; \lambda)=0$. The population penalty function $f_{\text{N}}(N(t); \lambda)$ of the Geometric Model seems unwieldy, but it does not require definition of additional parameters $\lambda$ and is in fact a monotonically non-decreasing function of $N(t)$ between $0 \leq N(t) \leq K$. Also note that $\lim_{N \to 0^+}f_{\text{N}}(N(t); \lambda) = 0$ and $\lim_{N \to K}f_{\text{N}}(N(t); \lambda) \to +\infty$, and this latter limit does not cause the Geometric Model to be undefined at $N(t)=K$ since overall $\lim_{N \to K} \dd N(t)/\dd t = 0$ due to the crowding function. Finally, Equation \eqref{eq:model_3} implies that $r$ can be physically interpreted as the per-capita growth rate when $N(t)$ is small and $E = 2E_{\text{c}}$, i.e.\ when the population is receiving double its minimum requirements of energy.

\paragraph{Steady states of the Geometric Model}\mbox{}\smallskip

The steady state populations for the Geometric Model, which are the equilibria of Equation \eqref{eq:model_3}, satisfy:
\begin{equation}
0 = r N^* \left( 1- \frac{N^*}{K} \right) \left[ \frac{E-E_{\text{c}}}{E_{\text{c}}} - \frac{K \ln\left((1-N^*/K)^{-1}\right) - N^*}{N^*} \right], 
\end{equation}
for which it is clear that $N^* = 0$, $N^* = K$, and there is a third steady state $N^*$ that satisfies the equation:
\begin{equation}
\frac{E-E_{\text{c}}}{E_{\text{c}}} = \frac{K \ln\left((1-N^*/K)^{-1}\right) - N^*}{N^*}.\label{eq:model_3_ss}
\end{equation}
Equation \eqref{eq:model_3_ss} has the solution:
\begin{equation}
N^* = \frac{K}{E} \left( E_c \, W \left( - \frac{E \exp(-E/E_c)}{E_c} \right) + E \right), \label{eq:model_3_lambert}
\end{equation} 
where $W(\cdot)$ is the principal branch of Lambert W function \citep{Corless1996}; the Lambert W function $W(x)$ outputs solutions $w$ of the equation $we^w = x$.

\begin{figure}[ht]
	\centering
		\includegraphics[width = 0.6\textwidth]{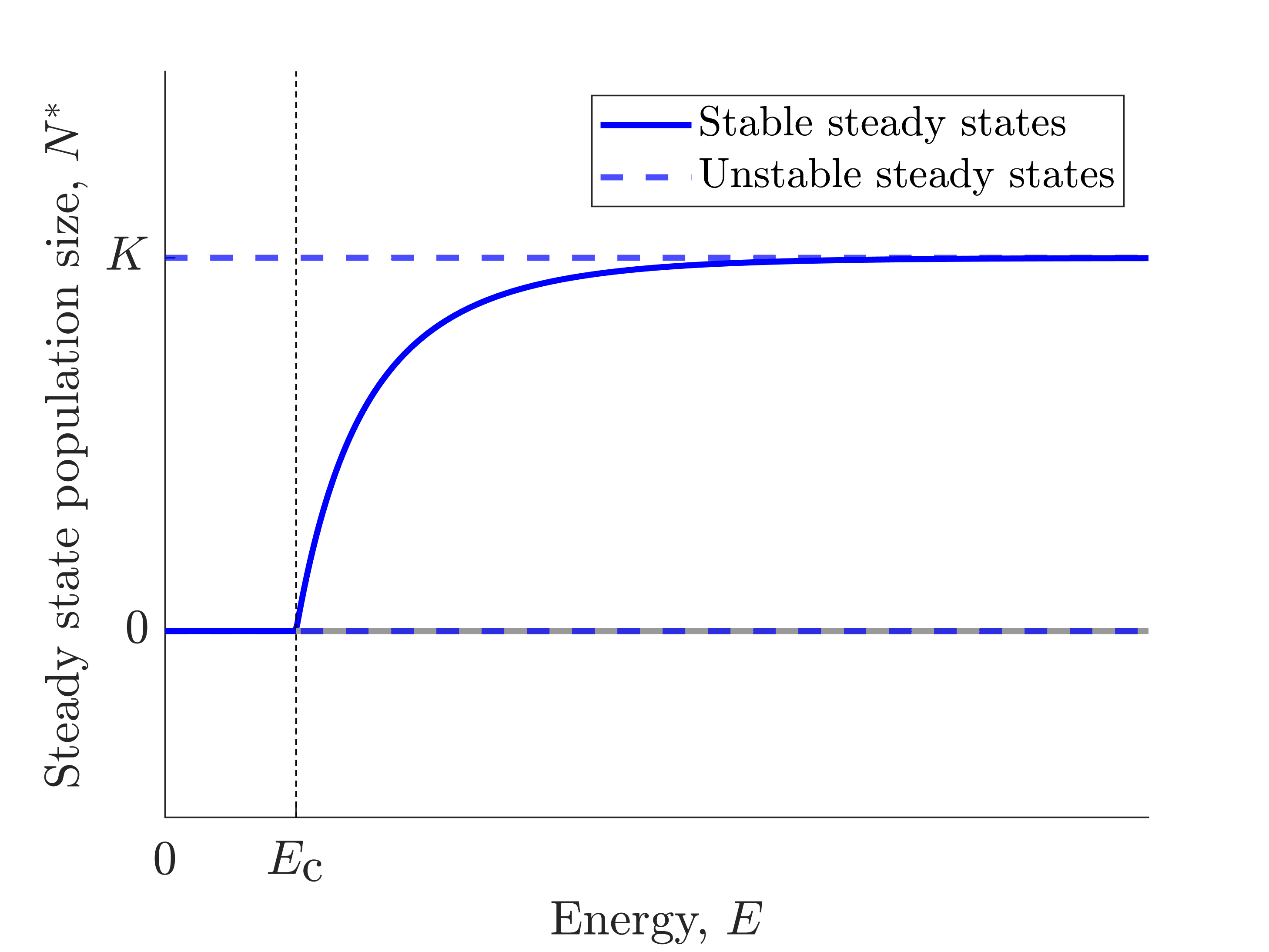}
		\caption{Steady states of the energy-dependent sigmoid growth model with logistic growth, linear energy dependence and nonlinear population penalty, derived from geometric arguments for plant cover vs light availability (Geometric Model, defined in Equation \eqref{eq:model_3}).}
		\label{fig:model_3_ss_all}
\end{figure}

The steady state $N^* = 0$ is stable for $E<E_{\text{c}}$ and the steady state represented by Equation \eqref{eq:model_3_lambert} is stable for $E>E_{\text{c}}$. The steady state $N^* = K$ is unstable. These steady states are depicted in Figure \ref{fig:model_3_ss_all}. A summary of the stable steady states for Models 1 to 3 is provided in Table \ref{tab:comparison_ss}.

\begin{table}[ht]
    \caption{Summary of stable steady states for the energy-dependent sigmoid growth models defined here as Models 1 to 3 (Equations \eqref{eq:model_1}, \eqref{eq:model_2} and \eqref{eq:model_3}, respectively). For a graphical description of the stable and unstable steady states, see Figures \ref{fig:model_1_ss}, \ref{fig:model2_ss_analysis_dim} and \ref{fig:model_3_ss_all}. $W(\cdot)$ is the principal branch of the Lambert W function \citep{Corless1996}.}
    \label{tab:comparison_ss}
    \centering
    \begin{tabular}{ c c c c } 
    \toprule
    Value of energy $E$ & Step Model & Linear Model & Geometric Model  \\ 
    \midrule
    $E < E_{\text{c}}$ & $0$ & $0$ & $0$ \\  [2\jot]
    $E_{\text{c}} < E < E_{\text{sat}}$ & $K$ & $K(E-E_{\text{c}})/(E_{\text{sat}} - E_{\text{c}})$ & $ \left(K/E)( E_c \, W \left( -E \exp (-E/E_{\text{c}}) /E_{\text{c}} \right) + E\right) $  \\ [2\jot]
     $E > E_{\text{sat}} $ & $K$ & $K$ & $ \left(K/E) (E_c \, W \left( -E \exp (-E/E_{\text{c}}) /E_{\text{c}} \right) + E\right) $\\
    \bottomrule
    \end{tabular}
\end{table}

\FloatBarrier

Advantageously, the Geometric Model only has three parameters to be estimated ($r$, $K$ and $E_{\text{c}}$), as it is a model based primarily on geometric arguments, and so allows for nuanced effects from environmental conditions in the form of energy or light received, $E$. For example, received light is a critical value for determining the composition of shallow water communities \citep{Miller2008}, and in particular the growth of macroalgae \citep{Dummermuth2003}. In the next section, we compare all three models described thus far for their ability to fit observed data for macroalgae populations constrained by light availability in shallow waters adjacent to Antarctica.

\subsection{Application to Antarctic benthic algae}

\subsubsection{Data sources and interpretation}\label{sec:data_sources}

\begin{figure}[ht]
	\centering
    \includegraphics[width = 0.5\linewidth]	{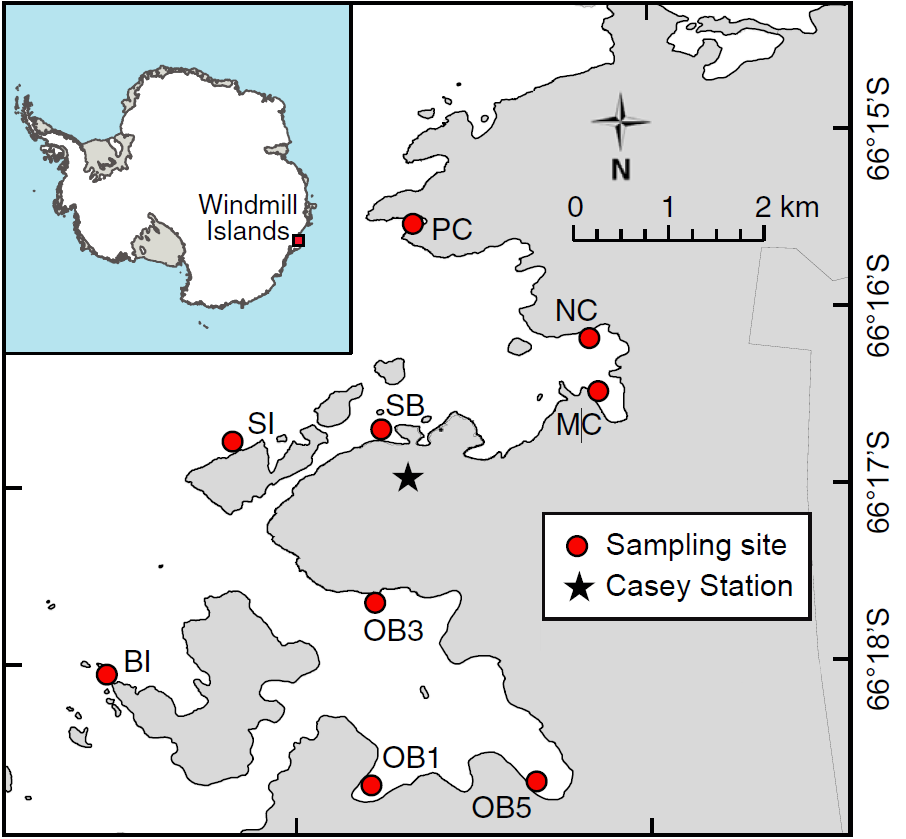}
    \caption{Map of study area where data used in the present study was collected. Inset shows the location of the Windmill Islands, on the coast of Wilkes Land, East Antarctica; all sampling sites were at this location. PC = Powell Cove, NC = Newcomb Corner, MC = McGrady Cove, SB = Shannon Bay, SI = Shirley Island, OB1 = O'Brien Bay 1, OB3 = O'Brien Bay 3, OB5 = O'Brien Bay 5 and BI = Beall Island. Sites PC and BI were excluded from the present study as there is no light data available for these sites. Figure adapted from \cite{Clark2013}.}  
    \label{fig:map_and_bars_good}
\end{figure}

Datasets were obtained by \cite{Clark2013} for several Antarctic shallow-water sites around Casey Station, located in the Windmill Islands, on the coast of Wilkes Land, East Antarctica (Figure \ref{fig:map_and_bars_good}). Of these sites, we used data from McGrady Cove (MC), Newcomb Corner (NC), O'Brien Bay 1 (OB1), O'Brien Bay 3 (OB3), O'Brien Bay 5 (OB5), Shannon Bay (SB) and Shirley Island (SI). The location of these sites are depicted in Figure \ref{fig:map_and_bars_good}. This data includes dates of sea-ice break-out and daily light measurements from each of the aforementioned sites \citep{CLARK2013data}. 

\FloatBarrier

\begin{figure}[ht!]
	\centering
    \includegraphics[width = 0.5\linewidth]	{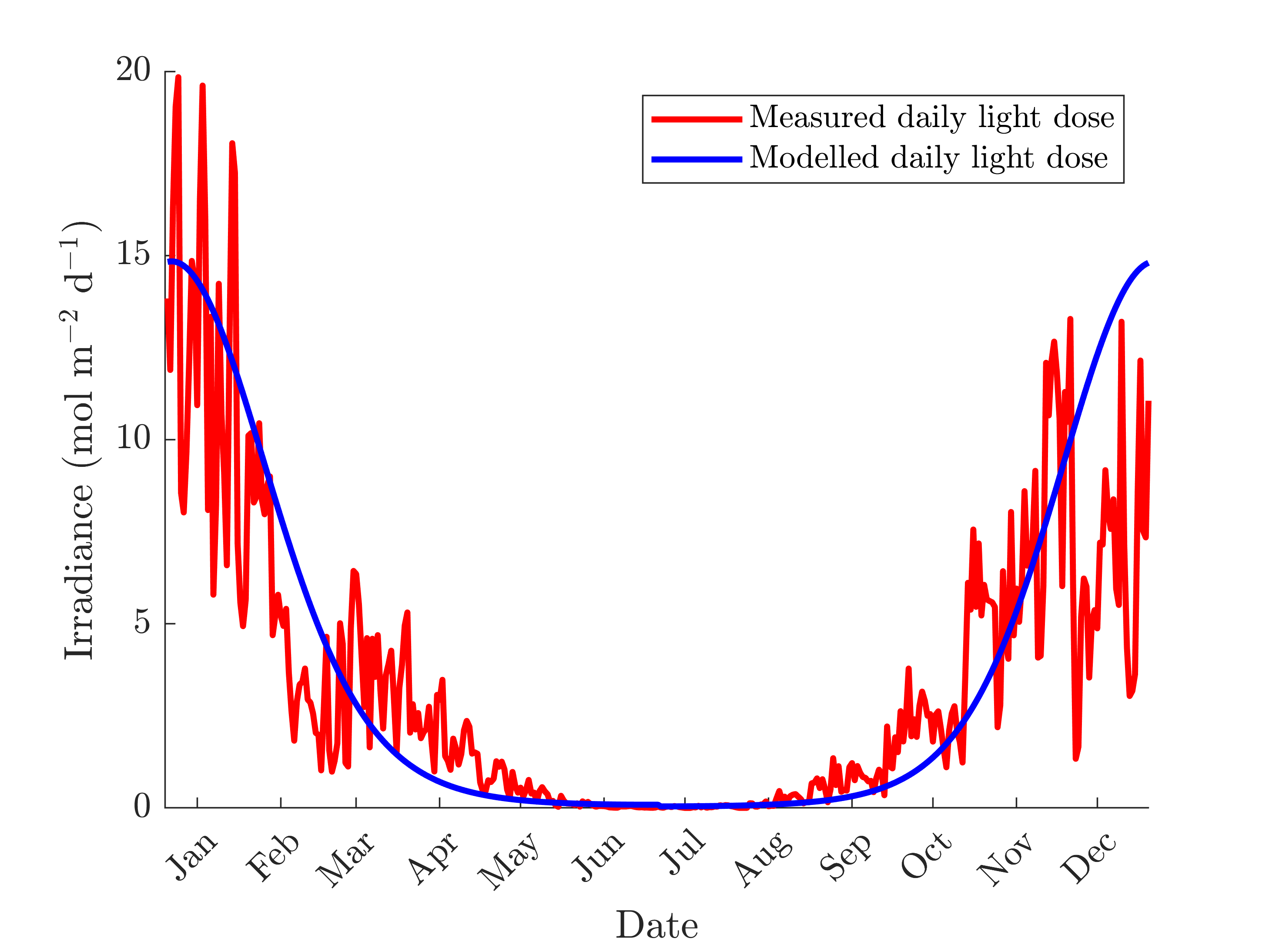}
    \caption{Comparison between the measured daily light dose (red) at the Shirley Island site (SI) and the daily modelled light dose as a periodic function of the date  by \cite{Clark2013} (blue).}  
    \label{fig:clark_modelled_light}
\end{figure}

Light measurements were obtained at each of the aforementioned sites using light sensors positioned at depths of 7 m to 10 m metres \citep{Clark2013}. Measurements from the Shirley Island site (SI) were used by \cite{Clark2013} to approximate the expected daily light reaching the sea-bed in the absence of sea-ice for one year: as sea-ice is only present at SI for approximately 3 months during winter when there is little daily light \citep{Clark2013}, SI is the site that most closely resembles conditions in the absence of sea-ice year-round. \cite{Clark2013} achieved this by fitting a periodic function to the light data from the SI site to obtain modelled daily light dose; a comparison between this modelled daily light dose and the measured daily light values is shown in Figure \ref{fig:clark_modelled_light}. The modelled daily light values (blue line in Figure \ref{fig:clark_modelled_light}) form the basis of our estimation of the expected light dose for different dates of sea-ice break-out (discussed in detail in Section \ref{sec:methods_light}). 

In this work we consider the \textit{daily} light dose, rather than \textit{instantaneous} light values, so all light measurements are stated in units of mol m$^{-2}$ d$^{-1}$ to allow direct comparison between the data collected by \cite{Clark2013} and the outputs of our models (discussed further in Section \ref{sec:results}). All light values used here refer to the solar radiation available for photosynthesis, commonly referred to as photosynthetically active radiation (PAR).

The data obtained by \cite{Clark2013} also includes multiple measurements of algae cover and invertebrate cover for each site, with measurements being taken at depths of 6 m and 12 m from both top-oriented and side-oriented surfaces of boulders. There are inherent differences in the amount of light available to algae on each of these boulder surface orientations, due to processes such as self-shading \citep{Baird2016, Baird2020}, sedimentation \citep{Knott2004, Clark2017} and light attenuation \citep{Mitchell1991} affecting these surfaces differently. 

Invertebrate cover measurements were reported by \cite{Clark2013} in two ways: cover occupied by all invertebrates, and cover occupied by all invertebrates excluding the two most abundant invertebrate species, \textit{Inversiula nutrix} and \textit{Spirorbis nordenskjoldi}, that were universally dominant at all sites. We will refer to these two measurement types as `abundant invertebrates' and `nonabundant invertebrates', respectively. 

Two algae species, \textit{Desmarestia menziesii} and \textit{Himantothallus grandifolius}, form algal canopies at the site SI. These canopies form at locations that have high light availability, and were therefore not observed at the other sites (Clark, pers.\ comm.). \cite{Clark2013} obtained data at the SI site from underneath both of these canopies, although we note the canopies themselves did not form part of the algae cover measurements. In the present work, in order to compare algae cover between sites, for SI we only used data taken from underneath canopies formed by \textit{Desmarestia menziesii}; this canopy is more diffuse and lets more light through than \textit{Himantothallus grandifolius}, and therefore more closely resembles the conditions of canopy absence (Clark, pers.\ comm.). 

There are three biological interpretations of the algae cover data in terms of the maximum cover that the algae can possibly occupy, which we will hereafter refer to as three different `interpretations' for ease of reference. Firstly, we can treat the algae cover measurements as absolute values, in which case we interpret that there is some non-biological and constant process responsible for preventing the maximum cover from reaching 100\% (e.g.\ sedimentation, which may affect measurements from the tops and sides of boulders differently) and therefore the carrying capacity $K < 100\%$. Secondly, we can consider the ratio of algae cover to the cumulative cover of algae and invertebrates, where we interpret any space not occupied by algae or invertebrates as being uninhabitable, with non-biological processes affecting each site differently, and therefore the carrying capacity $K = 100\%$. Finally, we can consider the ratio of algae cover to the cumulative cover of algae and nonabundant invertebrates, and so interpret that any space not occupied by algae or nonabundant invertebrates is uninhabitable, with the abundant invertebrate species and non-biological processes both limiting space and affecting each site differently, and therefore the carrying capacity $K = 100\%$. 

Data for the two boulder orientations (top and side) can be interpreted in any of the three options described above, resulting in what we refer to as six `scenarios'. Table \ref{tab:comparison_ss_full} summarises the algae cover data measured by \cite{Clark2013} for each scenario. We chose to aggregate the measurements taken at 6 m and 12 m depths to use in the present work, for three reasons. Firstly, although we expected a decrease in measured algae cover as the depth increased from 6 m to 12 m, this was only the case at two sites; one site consistently measured zero, while at the remaining four sites, for both top- and side-oriented surfaces, there was a clear \textit{increase} in measured algae cover as the depth increased from 6 m to 12 m. Secondly, we are operating under the assumption that the presence of sea-ice eliminates light completely, and that the presence or absence of sea-ice has a greater influence on light availability than the difference in depths. Finally, the light data was obtained at depths of 7 m to 10 m, which is not precisely representative of the depths of the algae measurements. The choice of aggregating the algae measurements at both depths will yield predicted measurements that most closely match with the light data available for the present study.

\begin{table}[H]
    \caption{Percentage of algae cover found at each site by \cite{Clark2013} (mean $\pm$ standard deviation, obtained from 16 replicate measurements) for each scenario (combination of boulder orientation and data interpretation). Sites are listed in order of decreasing annual light dose for each boulder surface orientation. Predicted annual light refers to an annual light dose calculated by combining a modelled curve fitted to Shirley Island's data (blue line in Figure \ref{fig:clark_modelled_light}) together with an assumed date of sea-ice break-out for each site (see Section \ref{sec:methods_light} for full details). See Figure \ref{fig:map_and_bars_good} caption for site abbreviations.}
    \label{tab:comparison_ss_full}
    \centering \scriptsize
    \begin{tabular}{ c c c c c c c } 
    \toprule
     Boulder & Site & Algae cover & Algae cover relative & Absolute algae & Measured & Predicted \\ 
     surface & & relative to all & to nonabundant &  cover (\%) & annual mean light & annual mean light\\
     orientation & & invertebrates (\%) &  invertebrates (\%) & & (mol m$^{-2}$ d$^{-1}$) & (mol m$^{-2}$ d$^{-1}$)  \\ 
    \midrule
    Side & SI & $41.0 \pm 28.4$ & $78.4 \pm 28.4$ & $31.9 \pm 25.0$ & 3.74 & 3.42 \\
    & OB3 & $60.1 \pm 19.7$ & $90.1 \pm 10.6$ & $15.1 \pm 13.6$ & 1.95 & 1.53 \\
    & NC & $23.7 \pm 30.6$ & $42.9 \pm 36.3$ & $6.7 \pm 10.3$ & 1.36 & 1.06 \\
    & SB & $19.5 \pm 21.1$ & $54.3 \pm 30.3$ & $6.4 \pm 7.9$ & 0.78 & 1.31 \\
    & MC & $10.8 \pm 12.9$ & $25.1 \pm 27.2$ & $4.0 \pm 5.3$ & 0.62 & 0.30 \\
    & OB5 & $17.8 \pm 24.8$ & $30.1 \pm 28.6$ & $8.1 \pm 13.4$ & 0.14 & 0.18 \\
    & OB1 & $0 \pm 0$ & $0 \pm 0$ & $0 \pm 0$ & 0.14 & 0.15 \\
    \midrule
    Top & SI & $48.3 \pm 37.5$ & $55.8 \pm 37.2$ & $46.1 \pm 36.9$ & 3.74 & 3.42 \\
    & OB3 & $91.2 \pm 7.0$ & $91.7 \pm 7.1$ & $46.0 \pm 19.1$ & 1.95 & 1.53   \\
    & NC & $40.1 \pm 34.9$ & $52.0 \pm 27.7$ & $20.0 \pm 21.5$ & 1.36 & 1.06  \\
    & SB & $75.3 \pm 24.3$ & $77.7 \pm 20.6$ & $36.2 \pm 32.2$ & 0.78 &  1.31 \\
    & MC & $35.8 \pm 23.6$ & $42.2 \pm 25.3$ & $16.9 \pm 18.8$ & 0.62 & 0.30  \\
    & OB5 & $52.0 \pm 37.8$ & $66.9 \pm 37.5$ & $44.4 \pm 36.4$ & 0.14 & 0.18 \\
    & OB1 & $0 \pm 0$ & $0 \pm 0$ & $0 \pm 0$ & 0.14 & 0.15 \\
    \bottomrule
    \end{tabular}
\end{table}

\subsubsection{Ranges of light values}\label{sec:methods_light}

In this work, we assumed that sea-ice reformation occurs on 1 April (as also assumed in \cite{Clark2013}) and therefore approximated the annual light dose for sea-ice break-out dates ranging from 2 April (1 day of ice cover) to 31 March the following year (365 days of ice cover). This was achieved using the modelled light function (blue line in Figure \ref{fig:clark_modelled_light}); we replaced the benthic light values modelled by \cite{Clark2013} with $E=0$ mol m$^{-2}$ day$^{-1}$ for days where sea-ice cover was assumed to be in place at each site (see Supplementary Material Appendix S1 to see how the measured and modelled curves compare) then obtaining the annual light dose by averaging over one year. This mathematical process assumes that the benthic light experienced by algae in the absence of ice cover follows the modelled curve shown in Figure \ref{fig:clark_modelled_light}.

\begin{figure}[ht]
	\centering
    \includegraphics[width = 0.7\linewidth]	{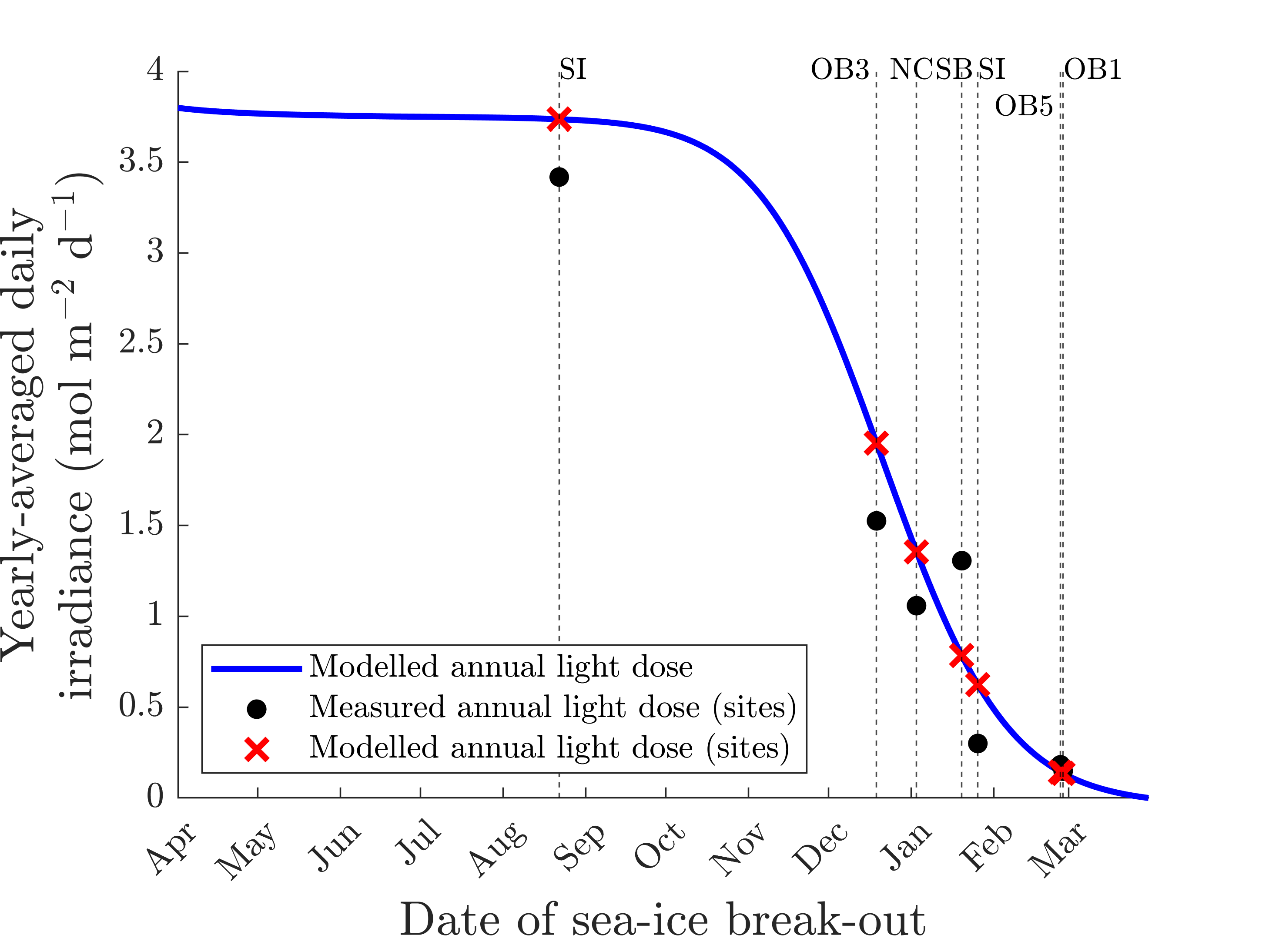}
    \caption{Modelled (blue) annual light dose, represented as daily light values averaged over one year, obtained by replacing benthic light values modelled by \cite{Clark2013} with $E = 0$ mol m$^{-2}$ d$^{-1}$ on days where the benthic communities are assumed to experience complete darkness due to sea-ice cover prior to the date of sea-ice break-out. Also shown is the measured annual light dose at each site (black) and the corresponding modelled annual light dose (red crosses). This figure assumes that sea-ice reforms on 1 April and the benthic light follows the modelled curve shown in Figure \ref{fig:clark_modelled_light}. See Figure \ref{fig:map_and_bars_good} for site abbreviations.}  
    \label{fig:meas_v_mod_light}
\end{figure}

Figure \ref{fig:meas_v_mod_light} shows the outcome of this mathematical process: the modelled annual light dose for each date of sea-ice break-out (blue line), as well as the measured (black circles) and modelled (red crosses) annual light dose at each of the seven sites. This light series is used in subsequent computations to estimate the timing of the tipping point between the benthic ecosystem being dominated by invertebrates or algae (discussed further in Section \ref{subsubsec:methods_tp}). 

\FloatBarrier

\subsubsection{Biological assumptions when fitting models to the data} \label{sec:biological_assumptions}

In the present analysis, we assumed that the algae populations have reached a steady state and are thus representative of habitat suitability. We recognise that this assumption, albeit implicitly made also in \cite{Clark2013}, is not guaranteed to hold \citep{Hastings2018}, but is practical given the available data. We consider the three models discussed in Sections \ref{sec:model_1} to \ref{sec:model_3} for their ability to fit the data from \cite{Clark2013}. In all three models, the growth rate parameter $r$ does not affect the steady state and therefore the data cannot be used to estimate this parameter, since we assume that algae populations have stabilised in time. However, if there are estimates of $r$ (such as in \citealt{Wiencke1990, Wiencke1990a, Goemez1997}), one could feasibility use these estimates together with the available data to parameterise the dynamic versions of the three models, if the desired application was to predict temporal changes in algae cover.

\subsubsection{Bayesian inference for fitting models to the data}\label{subsec:methods_bayesian}

The steady states of the three models detailed in Sections \ref{sec:model_1} to \ref{sec:model_3} (summarised in Table \ref{tab:comparison_ss}) were calibrated to each of the six `scenarios' (see Section \ref{sec:data_sources}) using Bayesian inference. Bayesian inference is a robust statistical framework that can be used for estimation and quantification of the uncertainty in model parameters \citep{Girolami2008}. Bayesian inference was implemented in this work using Sequential Monte Carlo (SMC) sampling due to its ability to exploit parallel computing architecture \citep{Dai2022}. 

Bayesian inference was used to estimate joint posterior distributions for between 2 and 4 parameters (depending on the scenario): the compensation irradiance $E_{\text{c}}$ (mol m$^{-2}$ d$^{-1}$), saturation irradiance $E_{\text{sat}}$ (mol m$^{-2}$ d$^{-1}$; estimated for the Linear Model only), carrying capacity $K$ (\% cover; estimated for one of the three data interpretations, see Section \ref{sec:data_sources}), and the standard deviation $\sigma$ characterising the model-data fit (in units of \% cover). A Gaussian likelihood function (characterised by the aforementioned standard deviation $\sigma$) was assumed. We are being agnostic about the source of this noise; it could be used to represent measurement uncertainty, how the ecology responds to the local environment, or a combination of both. A joint prior distribution was chosen to enforce physical requirements of the system, whilst simultaneously being as uninformative as possible \citep{Banner2020}. With the exception of the Linear Model (discussed in the next paragraph), the joint prior distribution was simply the product of independent uniform prior distributions for each parameter.

Although all models required estimation of the compensation irradiance $E_{\text{c}}$, the Linear Model was the only model we investigated that also required estimation of the saturation irradiance $E_{\text{sat}}$; since it is a physical requirement that $E_{\text{sat}} > E_{\text{c}}$ we enforced this in the prior distribution for model-data calibrations involving the Linear Model. This means that the joint prior distribution for the Linear Model parameters was only nonzero if $E_{\text{sat}} > E_{\text{c}}$, thus making the prior distribution for these two parameters non-independent in the Linear Model.

The lower and upper bounds of the prior uniform distributions for individual parameters were chosen as follows. For the standard deviation $\sigma$ and carrying capacity $K$ (where applicable), we chose prior bounds of [0, 100\%], representing the entire range of possible values for these parameters. For the compensation irradiance $E_{\text{c}}$, we chose prior bounds of [0, 5 mol m$^{-2}$ d$^{-1}$], because the minimum light requirements of algae should exceed zero light, and the chosen maximum bound for $E_{\text{c}}$ is greater than both the measured and modelled maximum annual light dose experienced by any site in our available data. For the saturation irradiance $E_{\text{sat}}$ we chose prior bounds of [0, 15 mol m$^{-2}$ d$^{-1}$], to ensure a sufficiently (but not excessively) wide range of possible values for this parameter. 

The SMC algorithm we implemented in this work was adapted from \cite{Adams2020a,Adams2020b}, with an additional resampling step included as described in \cite{Vilas2021} to ensure equal weighting of posterior samples. The SMC tuning parameters used in this work included the fraction of particles that must jump at each iteration, $C = 95$\%, and the effective sample size reduction target, $\Delta = 0.005$. For more information about what these tuning parameters represent, see Appendix D of \cite{Adams2020a}. We report the estimated value for each parameter as the mean $\pm$ standard deviation of the posterior samples (obtained from 1000 samples). To ensure that this number of samples was representative of the posterior distribution, all SMC simulations were run three times to confirm this number of samples was sufficient for reproducible results (not shown for brevity). Each calibrated model was then used to forecast algae cover over a range of annual light doses (discussed further in Sections \ref{sec:methods_predictions} and \ref{subsubsec:methods_tp}). Forecasts included 68\% and 95\% central credible intervals to represent uncertainty, calculated using the posterior predictive distributions obtained from the SMC sampling. 

\subsubsection{Model selection and prediction methods}\label{sec:methods_predictions}

When fitting each of the three models to the data of \cite{Clark2013}, the model with the lowest estimated value of the model-data fit discrepancy $\sigma$ was treated as the `best' model. We report these values of $\sigma$ in units of \% cover, and it is therefore an absolute rather than relative measure. This measure of goodness of fit is applicable only when model complexity is disregarded, and we recognise that formal model selection procedures could also be used to penalise model complexity if desired \citep{Johnson2004, Tredennick2021, Warne2019}.

\subsubsection{Prediction of tipping points}\label{subsubsec:methods_tp}

The calibrated models were used to forecast algae cover over a range of annual light doses and the corresponding range of sea-ice break-out timing dates (discussed in Section \ref{sec:methods_light}). These forecasts were used to estimate the date of sea-ice break-out at which a shift in the \textit{timing} of this date would yield the greatest change in steady state algae cover, as a coarse estimate for the timing of a tipping point in algae cover. We also calculated the predicted rate of change in algae cover with respect to the date of sea-ice break-out associated with each estimated tipping point. We report estimations of the greatest change in algae cover only for the Geometric Model as the mean $\pm$ standard deviation calculated from the posterior samples. In the case of the Step Model, the maximum rate of change is mathematically infinite at the location of the switch in steady states (see e.g.\ Figure \ref{fig:model_1_ss}), while in the case of the Linear Model, the tipping point predictions are a result of mathematical artefacts due to its functional form (for further information, refer to Supplementary Material Appendix S2). 

\section{Results}\label{sec:results}

\subsection{Fitting the models to the data}

All three tested models (described in Sections \ref{sec:model_1} to \ref{sec:model_3}) were fitted successfully to the six different datasets of algae cover versus annual light dose (termed `scenarios', see Section \ref{sec:data_sources}), using Bayesian inference implemented via SMC sampling (Table \ref{tab:analysis_summary_params}). 

\begin{figure}[H]
\captionsetup{skip=0\baselineskip}
\captionsetup[subfigure]{skip=0pt, singlelinecheck=false}
	\centering
	\begin{subfigure}[b]{0.4\textwidth}
		\centering
		\caption{Step Model, algae cover vs light dose}\includegraphics[width = \textwidth]{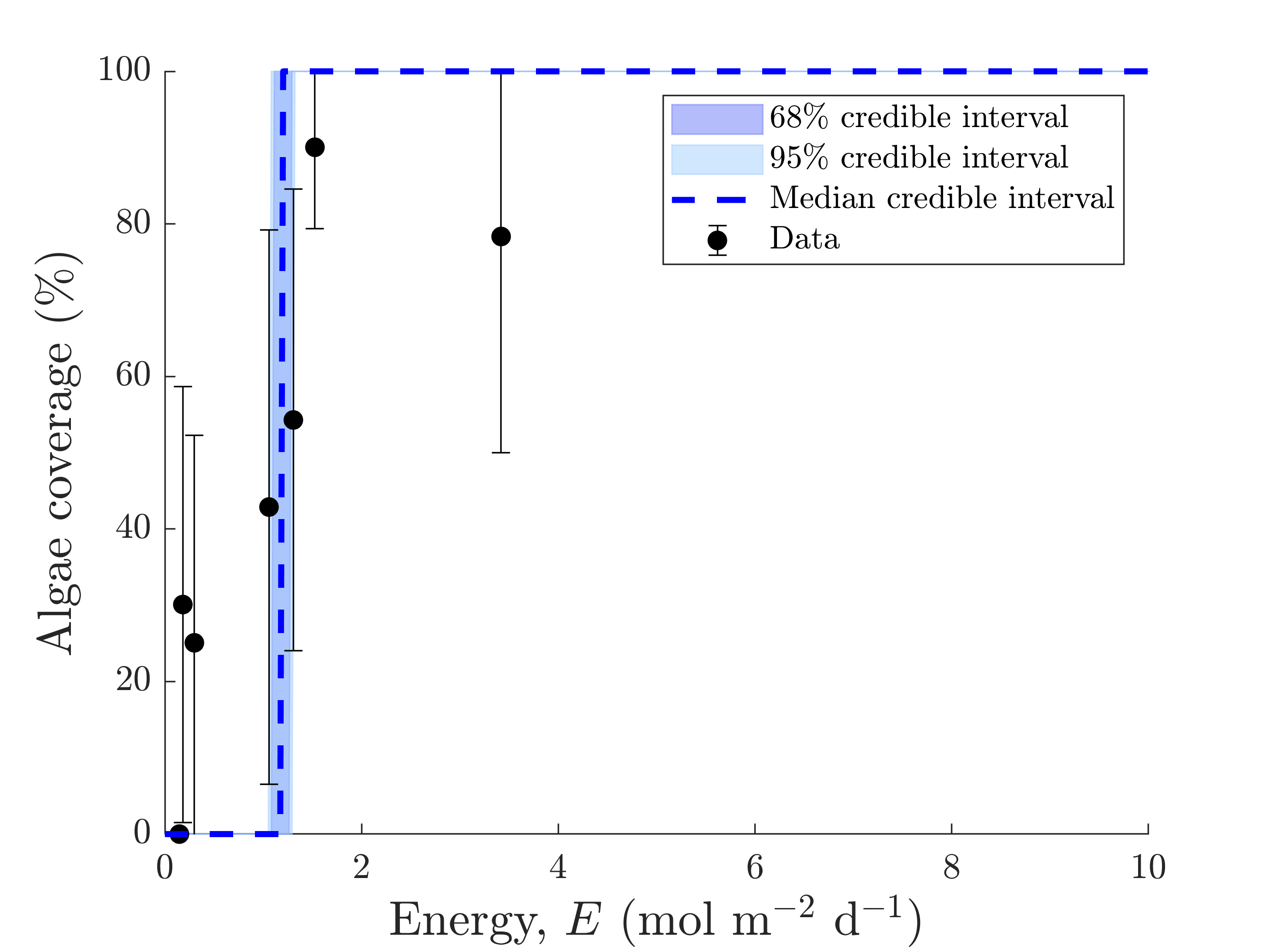}
	\label{fig:Model_1_A_v_E_sides_relative_to_nonabundant_invertebrates}
	\end{subfigure}
	\begin{subfigure}[b]{0.4\textwidth}
		\centering
		\caption{Step Model, algae cover vs sea-ice break-out date}\includegraphics[width = \textwidth]{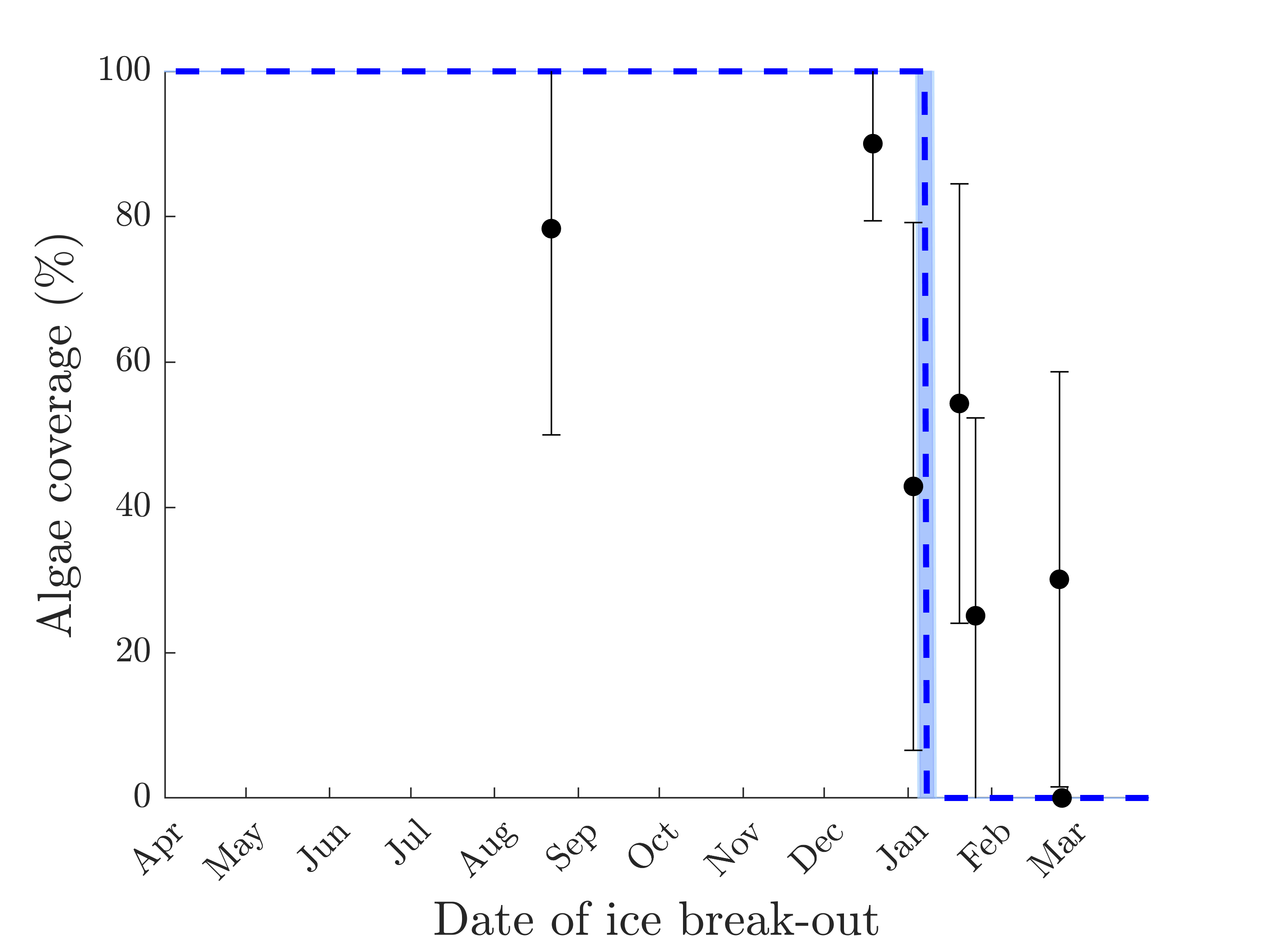}
		\label{fig:Model_1_A_v_melt_date_sides_relative_to_nonabundant_invertebrates}
	\end{subfigure}

	\begin{subfigure}[b]{0.4\textwidth}
		\centering
		\caption{Linear Model, algae cover vs light dose}\includegraphics[width = \textwidth]{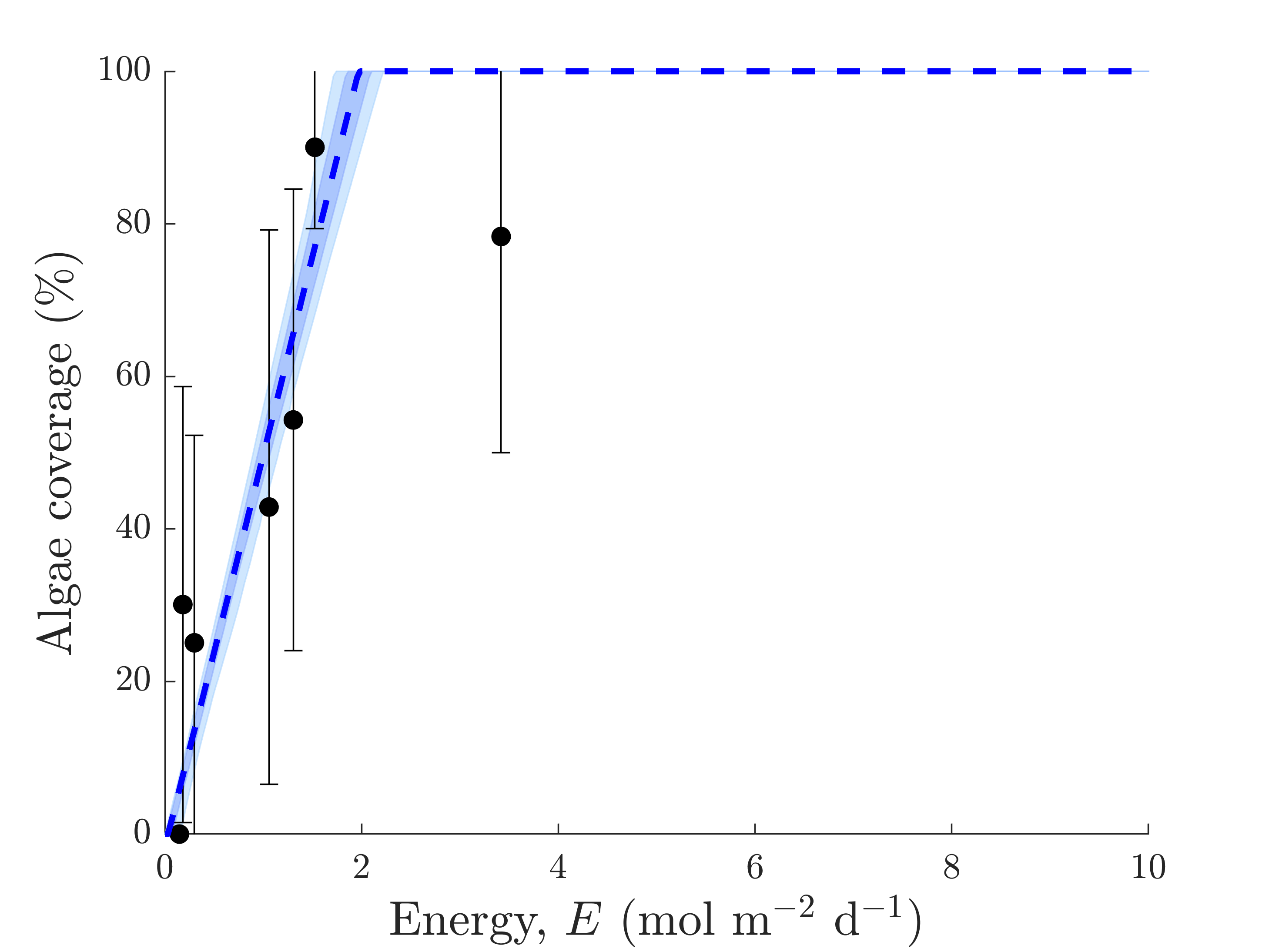}
		\label{fig:Model_2_A_v_E_sides_relative_to_nonabundant_invertebrates}
	\end{subfigure}
	\begin{subfigure}[b]{0.4\textwidth}
		\centering
		\caption{Linear Model, algae cover vs sea-ice break-out date}\includegraphics[width = \textwidth]{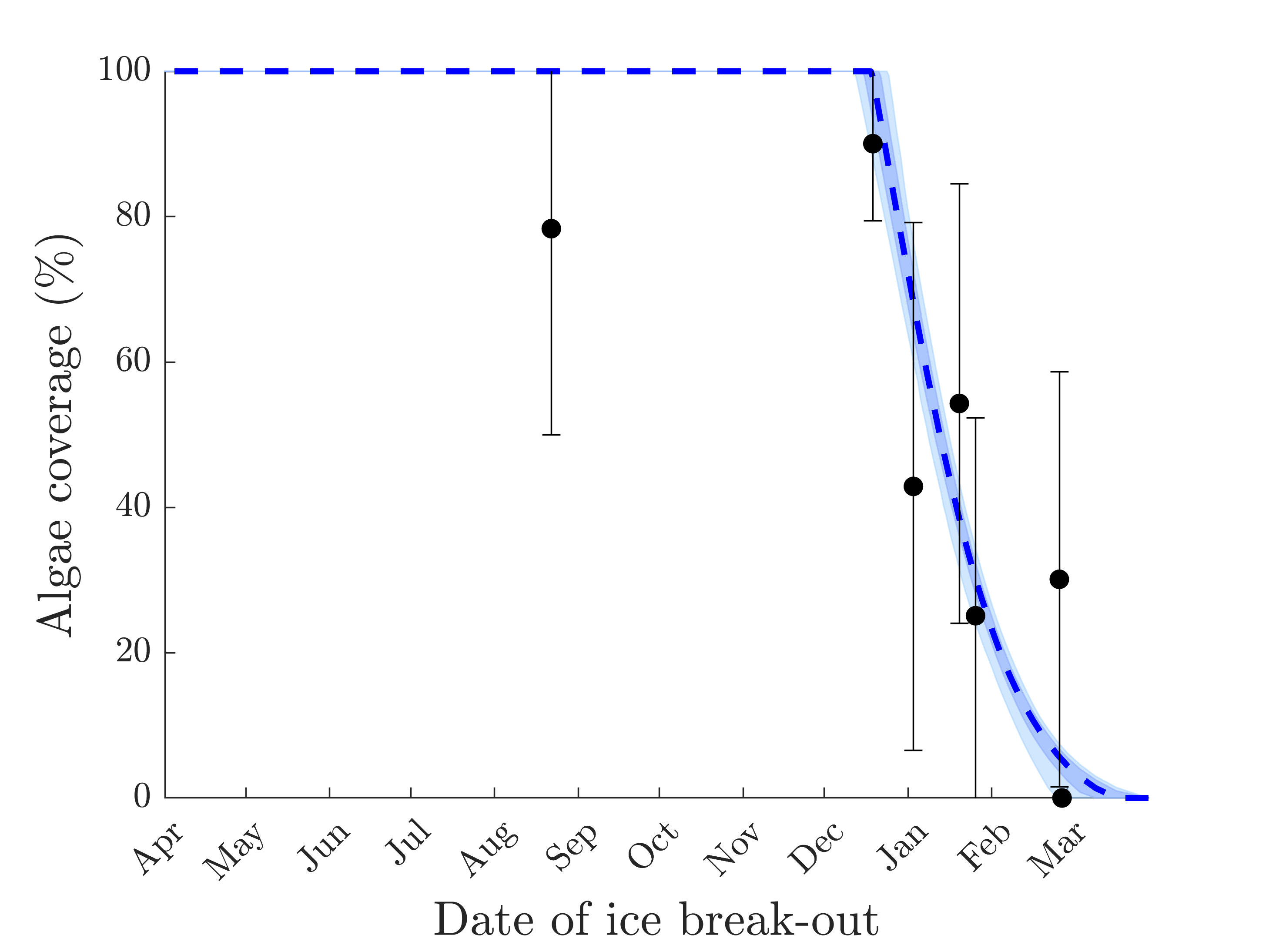}
			\label{fig:Model_2_A_v_melt_date_sides_relative_to_nonabundant_invertebrates}
	\end{subfigure}

	\begin{subfigure}[b]{0.4\textwidth}
		\centering
		\caption{Geometric Model, algae cover vs light dose}\includegraphics[width = \textwidth]{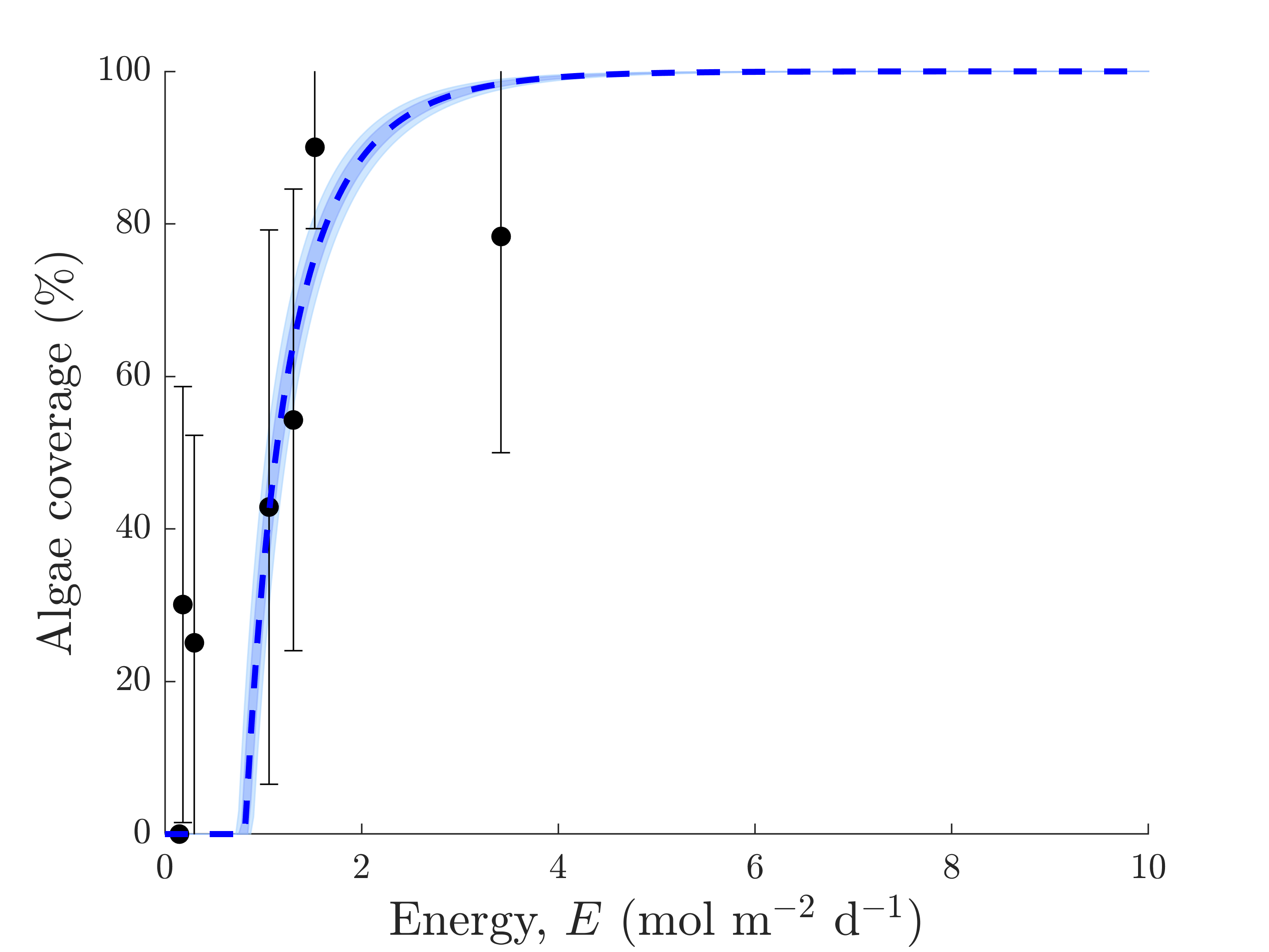}
				\label{fig:Model_3_A_v_E_sides_relative_to_nonabundant_invertebrates}
	\end{subfigure}
	\begin{subfigure}[b]{0.4\textwidth}
		\centering
		\caption{Geometric Model, algae cover vs sea-ice break-out date}\includegraphics[width = \textwidth]{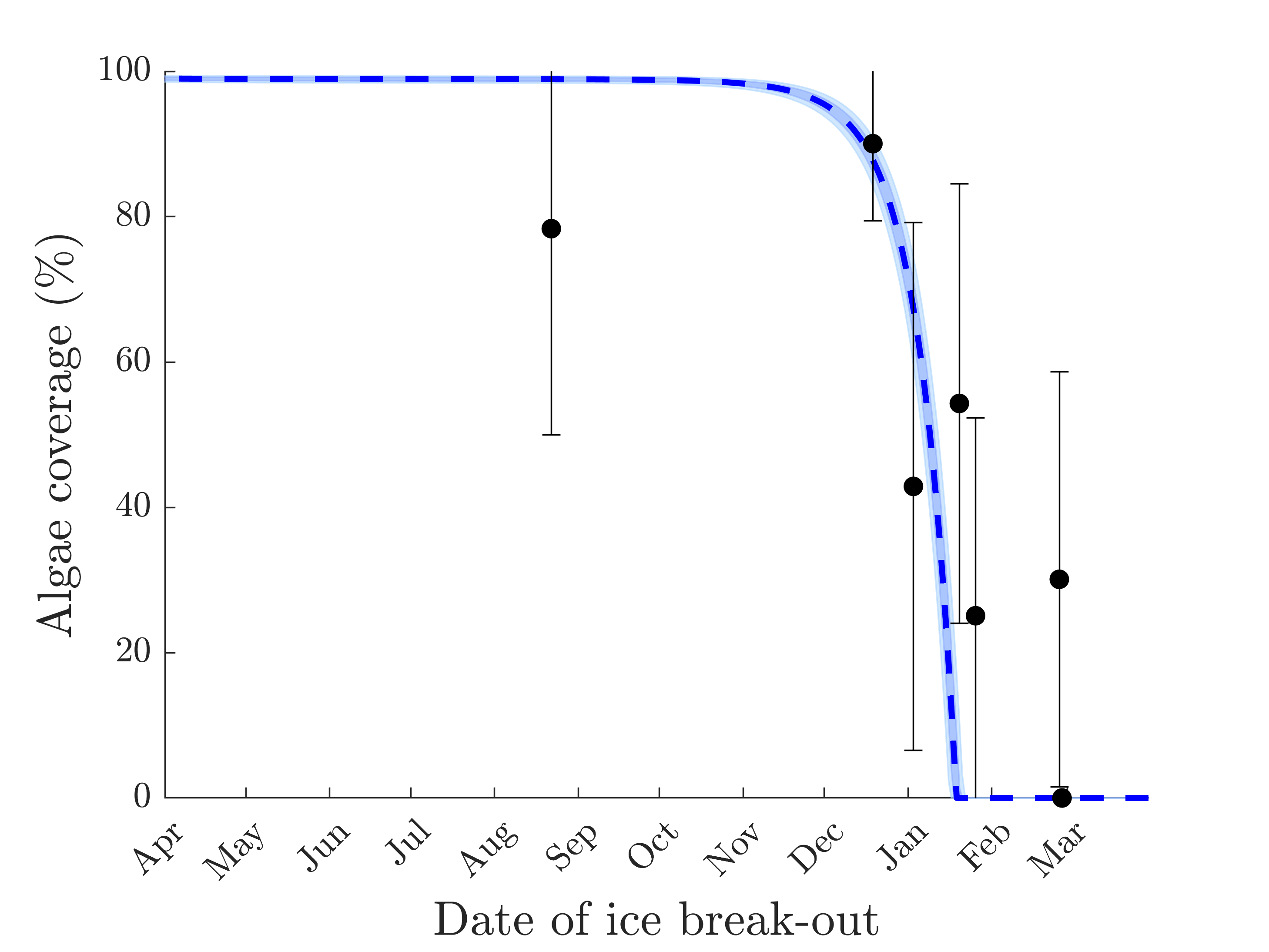}
				\label{fig:Model_3_A_v_melt_date_sides_relative_to_nonabundant_invertebrates}
	\end{subfigure}
	\caption{Model-data fits for side-oriented surfaces of boulders, interpreting the data as relative to the cumulative cover of nonabundant invertebrates and algae. Algae cover forecasts for a range of values of annual light dose $E$ (in units of mol m$^{-2}$ d$^{-1}$) are shown for the (a) Step Model, (c) Linear Model and (e) Geometric Model (see Sections \ref{sec:model_1} to \ref{sec:model_3} for model details). Plotting these forecasts against a range of dates of ice break-out for the (b) Step Model, (d) Linear Model and (f) Geometric Model demonstrate the nonlinear relationship between annual light dose and date of sea-ice break-out. Median model predictions are shown by the dashed lines, the 68\% credible intervals are shown by the dark blue shaded regions, and the 95\% credible intervals are shown by the light blue shaded regions. Note that these credible intervals represent uncertainty in best-fit models rather than uncertainty in model predictions of new observations, so the plotted credible intervals will necessarily be narrower than the data used to inform them. The forecasts are compared to the data obtained by \cite{Clark2013} at seven sites (Table \ref{tab:comparison_ss_full}). (Refer to Supplementary Material Appendix S3 for the model-data fits of the remaining five scenarios.)} 
	\label{fig:predictions_sides_relative_nonabundant}
\end{figure}

Figure \ref{fig:predictions_sides_relative_nonabundant} shows the model-data fits for one of the six `scenarios': data taken from the sides of boulders, interpreted as relative to the cumulative cover of algae and nonabundant invertebrates. Figures \ref{fig:Model_1_A_v_E_sides_relative_to_nonabundant_invertebrates}, \ref{fig:Model_2_A_v_E_sides_relative_to_nonabundant_invertebrates} and \ref{fig:Model_3_A_v_E_sides_relative_to_nonabundant_invertebrates} show how the models fit to the data for this particular scenario for a range of annual light doses, from 0 to 10 mol m$^{-2}$ d$^{-1}$. These figures show how the predicted algae cover is 0\% when the annual light dose $E < E_{\text{c}}$, and the algae cover increases towards the carrying capacity $K$ once the annual light dose exceeds $E_{\text{c}}$. Note that the plotted credible intervals represent uncertainty in the best-fit model rather than uncertainty in model predictions of new observations, so the plotted credible intervals will necessarily be narrower than the data used to inform them. Figures \ref{fig:Model_1_A_v_melt_date_sides_relative_to_nonabundant_invertebrates}, \ref{fig:Model_2_A_v_melt_date_sides_relative_to_nonabundant_invertebrates} and \ref{fig:Model_3_A_v_melt_date_sides_relative_to_nonabundant_invertebrates} show the algae cover model-data fits now plotted against the date of sea-ice break-out (for sea-ice cover ranging in duration from 1 to 365 days) instead of annual light dose. We note that there is a nonlinear relationship between the quantities of date of sea-ice break-out and annual light dose \citep{Clark2013}, and this is highlighted particularly when comparing Figures \ref{fig:Model_2_A_v_E_sides_relative_to_nonabundant_invertebrates} and \ref{fig:Model_2_A_v_melt_date_sides_relative_to_nonabundant_invertebrates} (the same algae cover model-data fits plotted against these two quantities).

\begin{table}[H]
    \caption{Predicted values for all model parameters, including the  compensation irradiance $E_{\text{c}}$, saturation irradiance $E_{\text{sat}}$, standard deviation $\sigma$ and maximum possible algae cover (carrying capacity $K$) (mean $\pm$ standard deviation, obtained from 1000 simulations) for calibrated models. Details of the models are provided in Sections \ref{sec:model_1} to \ref{sec:model_3}, and explanations of the data interpretations are provided in Section \ref{sec:data_sources}.}
    \label{tab:analysis_summary_params}
    \centering
    \begin{tabular}{ l c c c c c } 
    \toprule
    Scenario & Model & $E_{\text{c}}$   & $E_{\text{sat}}$ & $\sigma$ (\%) & $K (\%)$  \\ 
     & & (mol m$^{-2}$ day$^{-1}$) & (mol m$^{-2}$ day$^{-1}$) &  &  \\
    \midrule
    Top data, & Step & $1.18 \pm 0.08$ & - & $45.1 \pm 3.1$ & - \\
    relative to all & Linear & $0.04 \pm 0.07$ & $1.79 \pm 0.14$ & $38.9 \pm 2.7$ & - \\
     & Geometric & $0.77 \pm 0.06$ & - & $41.4 \pm 2.7$ & - \\
    \midrule
    Top data, & Step & $0.17 \pm 0.08$ & - & $44.4 \pm 3.0$ & -  \\
    relative to & Linear & $0.02 \pm 0.02$ & $1.64 \pm 0.19$ & $39.2 \pm 2.7$ & - \\
     nonabundant & Geometric & $0.17 \pm 0.13$ & - & $41.4 \pm 3.0$ & -  \\
    \midrule
    Top data,  & Step & $0.17 \pm 0.01$ & - & $28.4 \pm 2.1$ & $35 \pm 3$  \\
    absolute& Linear & $0.14 \pm 0.04$ & $0.22 \pm 0.22$ & $28.7 \pm 2.0$ & $35 \pm 3$  \\
     & Geometric & $0.12 \pm 0.04$ & - & $29.8 \pm 2.1$ & $36 \pm 3$ \\
    \midrule
    Side data, & Step & $3.31 \pm 1.36$ & - & $37.9 \pm 2.4$ & -  \\
    relative to all & Linear & $0.06 \pm 0.06$ & $6.02 \pm 0.60$ & $27.1 \pm 1.9$ & -  \\
     & Geometric & $1.13 \pm 0.10$ & - & $32.6 \pm 2.1$ & - \\
    \midrule
    Side data, & Step & $1.18 \pm 0.08$ & - & $39.2 \pm 2.7$ & -  \\
    relative to & Linear & $0.05 \pm 0.09$ & $1.97 \pm 0.13$ & $29.8 \pm 2.1$ & -  \\
    nonabundant & Geometric & $0.82 \pm 0.04$ & - & $31.4 \pm 2.1$ & - \\
    \midrule
    Side data,  & Step & $1.77 \pm 0.58$ & - & $14.7 \pm 1.1$ & $27 \pm 5$  \\
    absolute& Linear & $0.33 \pm 0.26$ & $7.24 \pm 2.33$ & $13.4 \pm 1.0$ & $69 \pm 20$ \\
     & Geometric & $1.14 \pm 0.09$ & - & $13.6 \pm 1.0$ & $33 \pm 4$ \\
    \bottomrule
    \end{tabular}
\end{table} 

\FloatBarrier

We next discuss, in the following sections, how our results show differences between the performance of each model in terms of fitting to each set of data (Section \ref{sec:results_model_fits}), differences between datasets in the estimated parameter values (Section \ref{sec:results_parameters}), and differences between datasets in both the predicted timing of tipping points and the rate at which algae cover increases at those tipping points (Section \ref{sec:results_tp}).

\subsection{Model fit comparison} \label{sec:results_model_fits}

\begin{figure}[h!]
	\centering
		\includegraphics[width = 0.6\textwidth]{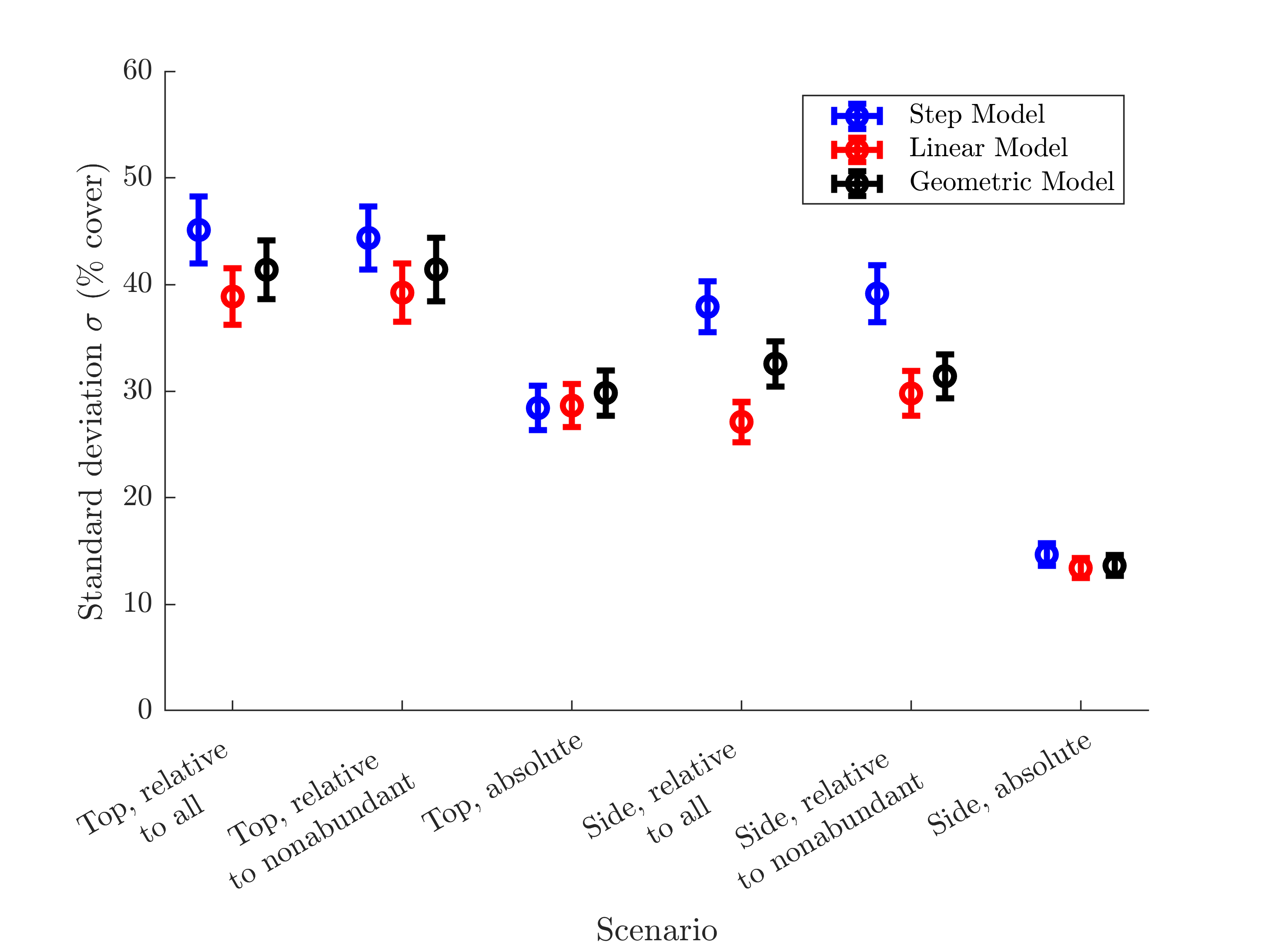}
		\caption{Estimates of the standard deviation $\sigma$ (mean $\pm$ error) which characterises the goodness-of-fit for all scenarios for each model. It is clearly seen that the estimates for the scenarios involving the side-oriented surfaces of boulders are lower than the scenarios involving the corresponding top-oriented surfaces, for all models.}
		\label{fig:summary_opt_sigma}
\end{figure}

The Linear Model was generally the best-fitting model (Table \ref{tab:analysis_summary_params}, Figure \ref{fig:summary_opt_sigma}), because it had the lowest value of $\sigma$ (see Section \ref{sec:methods_predictions} for justification of this metric) in five out of the six scenarios, followed by the Geometric Model. The Step Model generally had the highest values of $\sigma$. However, it is worth noting that the Linear Model has one additional free parameter ($E_{\text{sat}}$) that the other two models do not possess, which likely impacted this finding. Furthermore, the differences in $\sigma$ between datasets is larger than the differences in $\sigma$ between models, and there is a distinct improvement in model-data fit (reduced $\sigma$) from top- to side-oriented surfaces (Table \ref{tab:analysis_summary_params}, Figure \ref{fig:summary_opt_sigma}).

\subsection{Parameter estimations} \label{sec:results_parameters}

The vast majority of parameters estimated were well constrained by the data (see Supplementary Material Appendix S4 to see the marginal posterior distributions for all parameters). Thus, most parameters were practically identifiable \citep{Browning2020}, indicating the complexity of the model matched well with the available data in this study.

The compensation irradiance $E_{\text{c}}$ is an important parameter to estimate because it indicates the minimum annual light dose below which the habitat is unsuitable for long-term algae survival. Estimates of $E_{\text{c}}$ were highly variable, spanning two orders of magnitude from $\approx 0.02$ to $\approx 3$ mol m$^{-2}$ d$^{-1}$ (Table \ref{tab:analysis_summary_params}). 

\begin{figure}[h!]
	\centering
		\includegraphics[width = 0.6\textwidth]{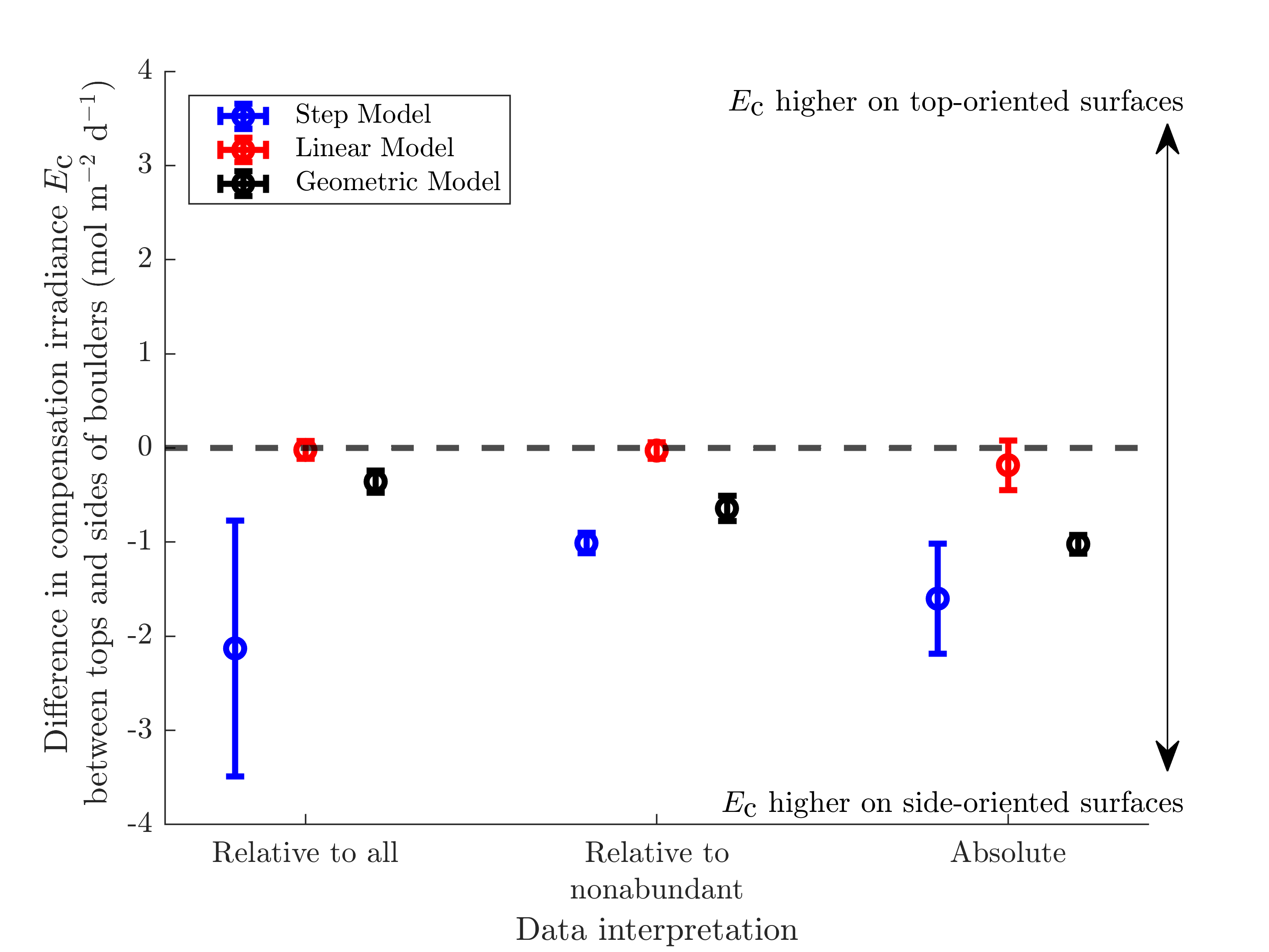}
		\caption{Difference between the model-predicted estimates of the compensation irradiance $E_{\text{c}}$ between tops and sides of boulders for each relevant combination of data interpretation and model. Details of the models are provided in Sections \ref{sec:model_1} to \ref{sec:model_3}, and explanations of the data interpretations are provided in Section \ref{sec:data_sources}.	Positive values indicate the value of $E_{\text{c}}$ is higher for top-oriented surfaces, while negative values indicate $E_{\text{c}}$ is higher for side-oriented surfaces. Error bars were calculated using standard error propagation formulae from the mean and standard deviations listed in Table \ref{tab:analysis_summary_params}.}
		\label{fig:difference_between_Ec}
\end{figure}

Interestingly, we found that for each model, the estimates for $E_{\text{c}}$ were always lower for the top-oriented surfaces, compared to the corresponding data taken from the side-oriented surfaces (Figure \ref{fig:difference_between_Ec}). It therefore follows that algae located on side-oriented surfaces require more light in order to achieve the same growth as algae located on top-oriented surfaces. This is physically reasonable because, for example, the effective area for algae to absorb light on side-oriented surfaces will be less than top-oriented surfaces as the former experience a greater impact from self-shading. 

For completeness, we acknowledge here that two of the 18 estimates of $E_{\text{c}}$, which are from the Step Model for the sides of boulders have high uncertainty (standard deviation $>0.5$ mol m$^{-2}$ d$^{-1}$, obtained from data interpreted as `relative to all' and `absolute', see Table \ref{tab:analysis_summary_params}). These uncertain estimates appear to be due to the limited number of sites at which sides of boulders were subjected to high annual light doses ($\gtrsim$ 2 mol m$^{-2}$ d$^{-1}$, see Figures S3.1 and S3.2).

The saturation irradiance $E_{\text{sat}}$, which indicates the maximum annual light dose above which the long-term algae cover cannot be further increased by receiving additional light, could be estimated from fitting the Linear Model to the data. These estimates of $E_{\text{sat}}$, like the estimates of $E_{\text{c}}$, were highly variable, ranging from $\approx 0.2$ to $\approx$ 7 mol m$^{-2}$ d$^{-1}$ (Table \ref{tab:analysis_summary_params}). Also similarly to $E_{\text{c}}$, estimates of $E_{\text{sat}}$ were always lower for the top-oriented surfaces, compared to data taken from the side-oriented surfaces. This provides further support for our (physically reasonable) assertion that algae located on side-oriented surfaces require more light than algae on top-oriented surfaces in order to maximise their growth. However, it should be noted that two of the six estimates of $E_{\text{sat}}$ exceeded the maximum measured annual light dose; additional data is therefore likely needed to provide a more definitive upper bound on the true value of $E_{\text{sat}}$. 

The maximum possible algae cover (carrying capacity $K$) was also predicted, for scenarios where the algae cover data was interpreted as absolute values. This maximum algae cover was predicted to be approximately 25\% to 35\% for five out of the six combinations of the three models (Step, Linear and Geometric) and two surface orientations (tops and sides of boulders). The sixth, outlier, prediction of carrying capacity $K$ was made using the Linear Model, which requires more data than the other two models to estimate $K$ due to its additional free parameter $E_{\text{sat}}$, on a dataset which is less informative about carrying capacity (Figure S3.1) than the other (`absolute') dataset tested (Figure S3.3).

Unlike the values predicted for $E_{\text{c}}$ and $E_{\text{sat}}$, where the magnitude of the predicted values were dependent on the boulder surface orientation, the values predicted for $K$ were more consistent, with no obvious difference between boulder surface orientations. We therefore hypothesise that 25\% to 35\% cover could represent an upper limit to how much space may be occupied by benthic algae in the shallow waters around Antarctica.

\subsection{Tipping point predictions}\label{sec:results_tp}

Possible tipping points were identified from the annual light dose, and corresponding sea-ice break-out date, that yielded the greatest predicted rate of change in algae cover, based on the fit of the Geometric Model to the data (the Linear and Step models are unsuitable for predicting the tipping point in this work; see Section \ref{subsubsec:methods_tp} for full technical details).

The maximum rate of change in algae cover (i.e.\ the rate of change in algae cover at the tipping point) was predicted to range between $\approx 2$ to $\approx 10$ \% cover per day, depending on the data used to inform this estimate (Table \ref{tab:analysis_summary_ROC_model3}). We note that the predicted maximum rate of change is lower for all side-oriented boulder surfaces than the corresponding top-oriented surfaces, consistent with previous results showing a difference between top- and side-oriented surfaces.

\begin{table}[H]
    \caption{Values predicted by the Geometric Model for the timing of the tipping point and value of the maximum rate of change in algae cover per day at this tipping point (mean $\pm$ standard deviation, obtained from 1000 posterior samples) for the calibrated model. Note we did not consider any tipping point predictions for the Step Model or Linear Model as discussed in Sections \ref{subsubsec:methods_tp}.}
    \label{tab:analysis_summary_ROC_model3}
    \centering
    \begin{tabular}{ l c c } 
    \toprule
    Scenario & Maximum rate of change  & Date   \\ 
     & (\% day$^{-1}$) &   \\
    \midrule
    Top data, relative to all & $6.8 \pm 0.2$ & 20 January $\pm$ 2 days  \\
    Top data, relative to nonabundant & $10.1 \pm 0.8$ & 23 February $\pm$ 8 days  \\
    Top data, absolute &  $3.9 \pm 0.4$ & 28 February $\pm$ 3 days  \\
    Side data, relative to all & $5.8 \pm 0.3$ & 8 January $\pm$ 3 days  \\
    Side data, relative to nonabundant &  $6.7 \pm 0.2$ & 18 January $\pm$ 1 day \\
    Side data,  absolute &  $1.9 \pm 0.2$ & 8 January $\pm$ 3 days \\
    \bottomrule
    \end{tabular}
\end{table}

\begin{figure}[h!]
	\centering
		\includegraphics[width = 0.8\textwidth]{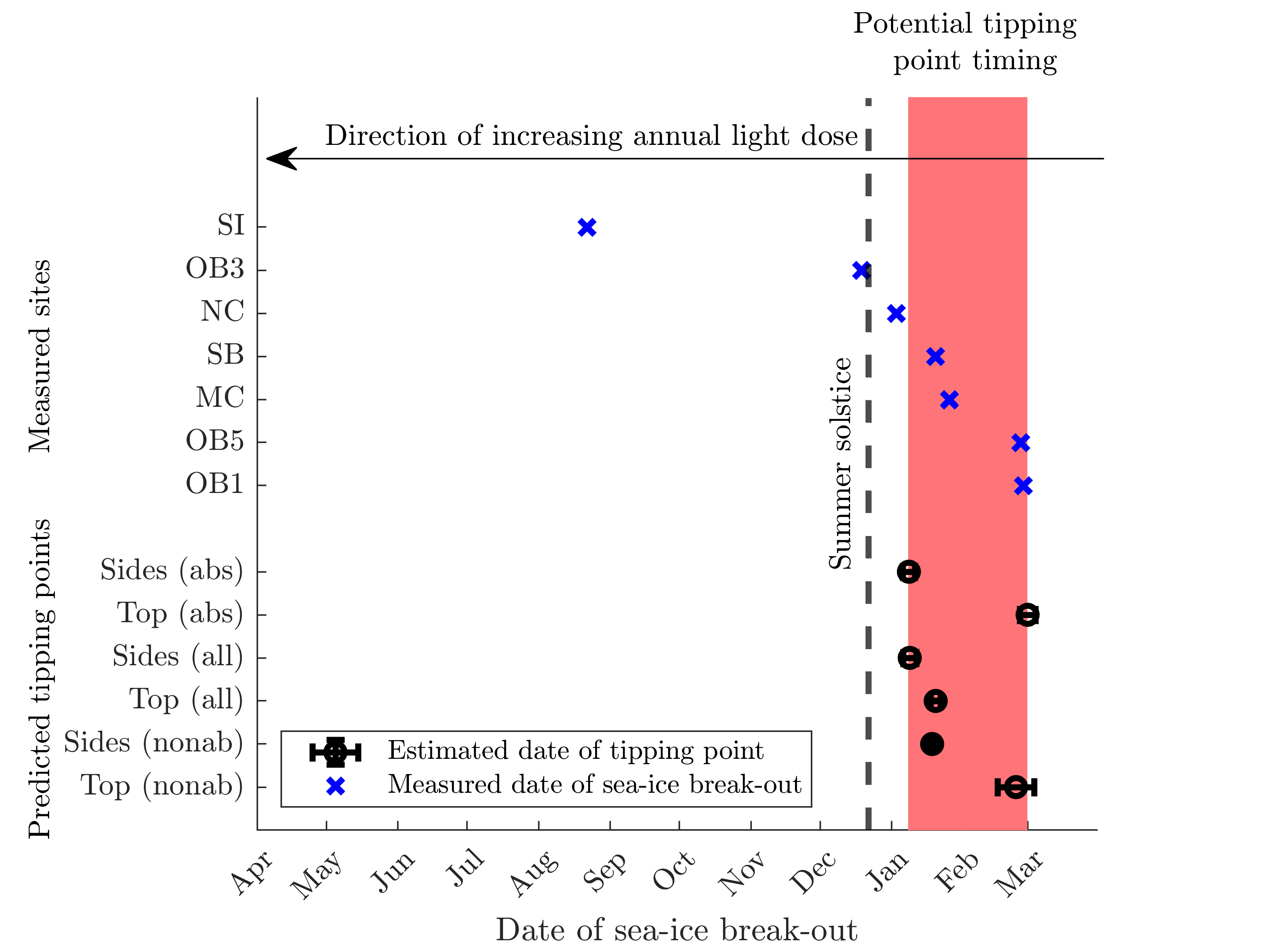}
		\caption{Comparison of measured sea-ice break-out dates for seven sites in the Windmill Islands against our predicted tipping point dates based on fitting energy-dependent sigmoid growth models to data from these sites. The blue crosses indicate the measured date of sea-ice break-out for each of the seven sites (discussed in Section \ref{sec:data_sources}). The black circles with error bars indicate the estimated date of sea-ice break-out ($\pm$ standard deviation) at which the maximum rate of change in algae cover is reached, as a proxy for the date of the tipping point. Note we did not consider any tipping point predictions for the Step Model or Linear Model as discussed in Sections \ref{subsubsec:methods_tp}. In this figure, as the period of sea-ice cover decreases, the date of sea-ice break-out moves earlier in the year (i.e.\ from right to left). The date of the summer solstice is shown by the black dashed line. The red shaded region indicates the range of dates of sea-ice break-out within which the tipping point was predicted by our models. Our results indicate that the seven sites are either in the vicinity of, or have already passed, the tipping point.}
		\label{fig:summary_date_of_max_ROC_model3}
\end{figure}

In four out of six scenarios, the Geometric Model predicted that the tipping point occurs in January, while in the remaining two scenarios, the tipping point was predicted to occur in February (Table \ref{tab:analysis_summary_ROC_model3}, Figure \ref{fig:summary_date_of_max_ROC_model3}). Interestingly, all predicted tipping points are predicted to fall later in the year than the summer solstice (compare red shaded region and black dashed line in Figure \ref{fig:summary_date_of_max_ROC_model3}). This means that for a decreasing period of sea-ice cover, where the date of sea-ice break-out moves earlier in the year, these dates will be reached \textit{before} the date of sea-ice break-out coincides with the summer solstice. Therefore, our work predicts that if the sea-ice breaks out earlier in the year than the summer solstice (to the left of the black dashed line in Figure \ref{fig:summary_date_of_max_ROC_model3}), the tipping point has already been passed, and the site is highly likely to be suitable for algae growth on both tops and sides of boulders.

\section{Discussion}\label{sec:discussion}

\subsection{Model framework}

In this work we have introduced and demonstrated a new model framework that can yield steady state populations with a nontrivial dependence on the energy available to the system, overcoming the issues with more traditional phenomenological modelling approaches described in Section \ref{sec:intro}. We have taken three model examples of this framework and shown how they fit with data from a case study system in Antarctica, confirming the models are capable of predicting key ecological parameters and tipping points, without the requirement for complex models with a large number of parameters.

While our work has focussed specifically on the model framework's application to the Antarctic case study, we note it is general in nature and has a broad application to any system requiring a dependence on (some form of) energy. Its construction allows incorporation of many existing sigmoid growth models (discussed by \citealt{Banks1993}) via modification of the crowding function $f_{\text{c}}(N(t);K,\lambda)$. The model framework's generality also makes it flexible; it can be as coarse or detailed as required, and can easily be adapted to suit specific systems if desired, via modifications of the framework's required functions $f_{\text{c}}(N(t);K,\lambda)$, $f_{\text{E}}(E;\lambda)$ and $f_{\text{N}}(N(t);\lambda)$.

However, we reiterate that this framework is only defined for a population $N(t)$ satisfying $0 \leq N(t) \leq K$. While this is common in many biological and ecological systems, our framework may require modifications if it is to be applied to systems where this inequality does not hold. Care must also be taken to ensure the selections for $f_{\text{c}}(N(t);K,\lambda)$, $f_{\text{E}}(E;\lambda)$ and $f_{\text{N}}(N(t);\lambda)$ avoid non-physical behaviour and adhere to the mathematical requirements set out in Section \ref{sec:growth_models}.

\subsection{Choosing models from the framework for Antarctic algae}\label{sec:disc_choosing_models}

Of the three example models examined in this work, our results found that the Linear Model was the most suited for forecasting Antarctic benthic algae cover. The Linear Model was the best-performing model when fitting to the available data. We therefore suggest that the Linear Model should be used for forecasting algae cover or determining the value of compensation irradiance $E_{\text{c}}$, which is a parameter that is biologically meaningful \citep{Adams2017} as a metric of minimum light requirements for algae. However, if the desired application was to determine information (e.g.\ the date) about potential tipping points, the Geometric Model is better suited (see Section \ref{subsubsec:methods_tp} for full details). 

However, regardless of each of the three models' ability to represent Antarctic benthic algae, these models all have general merits and are likely to find a broad application in many other ecological contexts. Although the Step Model was generally not suitable for forecasting benthic algae cover or quantifying tipping points, we note this model could be used to represent any system where limited data is available to characterise the shape of the dependence of steady state populations on energy and/or systems that do not have a justification for usage of the more complicated Geometric Model. The Linear Model could similarly be applied to any system where steady state populations possessing linear dependence on energy (within a finite range of energy values) is an ideal first model to trial (i.e.\ in the absence of other mechanistic understanding). On the other hand, the Geometric Model (due to its geometric derivation from plant cover as a function of available light) provides a mechanistic understanding of benthic plant growth or cover in response to light, so is well suited for such applications.

\subsection{Predicting the light requirements of Antarctic algae}\label{sec:disc_light_req}

We next turn our attention to the comparison of our data-calibrated models' predictions of key ecological parameters for Antarctic algae, starting with the compensation irradiance $E_{\text{c}}$. A range of photosynthetic parameters for different algal species were calculated experimentally by \cite{Clark2013}, including the minimum annual light balance (mALB) for four nonabundant algae species (Table \ref{tab:clark_params}). Focussing here on our best-fitting model (the Linear Model), which predicted values for the compensation irradiance $E_{\text{c}}$ ranging from $\approx 0.02$ to $\approx 0.33$ mol m$^{-2}$ d$^{-1}$ (Table \ref{tab:analysis_summary_params}), this range finds reasonable agreement with the mALB values of 0.074 to 0.690 mol m$^{-2}$ d$^{-1}$ experimentally determined by \cite{Clark2013}. In particular, we found the predicted values of $E_{\text{c}}$ from the Linear Model were more consistent with the minimum annual light dose for the algae species \textit{Desmarestia menziesii} (0.074 mol m$^{-2}$ d$^{-1}$). As our models do not distinguish between different algae species, this may mean that our data-calibrated Linear Model will predict algae growth to be more consistent with \textit{Desmarestia menziesii} than other more light-sensitive species.

\begin{table}[H]
    \caption{Minimum annual light balance experimentally determined by \cite{Clark2013} for the four nonabundant algae species. The equivalent daily light dose was obtained by calculating mALB/365. The resulting range between the minimum and maximum bounds is also shown.}
    \label{tab:clark_params}
    \centering
    \begin{tabular}{ l c c c } 
    \toprule
     Species & mALB & mALB/365    \\ 
     & (mol m$^{-2}$ yr$^{-1}$) & (mol m$^{-2}$ d$^{-1}$) \\
     \midrule
     \textit{Desmarestia menziesii} & 27 & 0.074 \\
    \textit{Himantothallus grandifolius} & 216 & 0.592 \\
     \textit{Iridaea cordata} & 127 & 0.348 \\
     \textit{Palmaria decipiens} & 252 & 0.690 \\
     \midrule
     Range & 27-252 & 0.074-0.690  \\
    \bottomrule
    \end{tabular}
\end{table}

Our results show that the estimated values for the compensation irradiance $E_{\text{c}}$ (and for the Linear Model, the saturation irradiance $E_{\text{sat}}$) are always lower for the top-oriented surfaces, compared to the corresponding side-oriented surfaces (Section \ref{sec:results_parameters}). Physically, these differences make sense. Geometrically, the orientation of side-facing surfaces compared to available light results in a smaller effective area (that is, the fraction of the area covered by algae as viewed from above) than top-facing surfaces due to self shading \citep{Baird2016,Baird2020}. As a result, algae on side-oriented surfaces will absorb less light than top-oriented surfaces for the same amount of available light. 

Light attenuation is another factor which may affect the amount or intensity of light delivered to different boulder surfaces \citep{Mitchell1991, Deregibus2016, Stambler1997}. Light attenuation can be different for different regions \citep{Deregibus2016}, but also for different depths in the water column \citep{Stambler1997}, whereby surfaces located at greater depths will receive less light. In the present work we were unable to explore the variations in the benthic community composition with depth due to the limitations of the data. We expect that depth could play a role in determining this composition, although the presence or absence of sea-ice likely remains the greatest control on light availability for Antarctic shallow-water benthic communities. 

Finally, the physical characteristics of the surrounding environment must also be considered. The data obtained by \cite{Clark2013} involves observations taken from various boulder surfaces; due to the irregular nature of such surroundings, it is therefore possible that, particularly for the side-oriented surfaces, the sample sites will be shaded (or partially shaded) by other boulders or structures. For side-oriented boulder surfaces, shading could even be caused by the boulder from which the sample was obtained, particularly in circumstances where the side-oriented surface was not strictly perpendicular to the seabed.

\subsection{Predicting the maximum cover, and rate of change in cover, of Antarctic algae}\label{sec:disc_max_cover}

In our results, we also found that the maximum algae cover possible, also known as the carrying capacity $K$, was predicted to be approximately 25\% to 35$\%$ (Section \ref{sec:results_parameters}). Recalling that $K$ was only estimated when interpreting the data as absolute values, and that for this data interpretation it is assumed there is a non-biological and constant process preventing the maximum cover from reaching 100$\%$, these results could represent a plausible upper limit for benthic algae growing in the shallow waters around Antarctica. Potential processes for limiting the maximum algae cover could include, for example, ice scour from bodies of ice colliding with the sea floor and/or substrate \citep{Barnes1999, Peck1999, Brown2004, Johnston2007, Smale2007, Barnes2011}, whereby continual collisions repeatedly clear fauna prior to it reaching full cover, although it is not clear how frequently this process occurs \citep{Barnes1999, Brown2004}.

Our analysis of tipping points suggested that there are some sea-ice break-out dates that can yield a change in steady state algae cover of up to $\approx 4$\% absolute cover or up to $\approx 10$\% relative cover if the ice breaks out one day earlier (Table \ref{tab:analysis_summary_ROC_model3}). It is important to note that these maximum rates of change in algae cover are a worst case scenario, although these predictions reflect how quickly the habitat suitability could change. If the maximum rate of $\approx 10$\% \textit{relative} algae cover per day is compared to expected maximum rates of sea-ice decline of up to 5 days per year estimated elsewhere \citep{Clark2013, Barnes2011, Parkinson2004, Stammerjohn2008}, these predictions together suggest that the habitat suitability for algae could possibly increase by $\approx 50$\% relative cover in a single year, if the maximum rates of sea-ice cover date and algae cover suitability (due to the tipping point) temporally coincide. Similarly, if the maximum rate of $\approx 4$\% \textit{absolute} algae cover per day were to coincide with the maximum rate of sea-ice decline of 5 days per year, this would suggest the habitat suitability could possibly increase from zero to near its maximum (25\% to 35\% absolute cover) in a single year. Hence, our results highlight the value of monitoring changes in the sea-ice break-out date, especially if this date is in the vicinity of the dates where the tipping point may occur (Table \ref{tab:analysis_summary_ROC_model3}), if warning of a potential change in benthic community dominance between algae and invertebrates in shallow coastal waters of Antarctica is of interest.

\subsection{Extending steady state models of Antarctic algae cover to dynamic models}\label{sec:dynamic_models}

We have made the assumption that the algae and invertebrate data collected by \cite{Clark2013} were from populations that had stabilised in time. This may not be the case \citep{Hastings2018}; however, without regular cover data measured repeatedly over a period of time, we cannot make cover predictions, or even compare the data between sites, unless we make this assumption. 

The models we introduce only require one additional parameter, the growth rate $r$, to be estimated so that the models become fully-parameterised dynamic models of algae cover. The inherent challenge with incorporating this parameter into the model is that these ecosystems are composed of unknown proportions of a varying number of algae species \citep{Clayton1994} of different sizes and at different stages in their life cycles. Each species has a different growth rate \citep{Johnston2007}, which itself varies both between male and female gametophytes, and across the year \citep{Wiencke1990}. However, regardless of these complexities, the fact that only one additional parameter (the growth rate $r$) is required to transform our fitted steady state models into dynamic models suggests this is a logical research direction for future work. 

\section{Conclusion}

In this work, we have demonstrated a newly proposed model framework with an explicit resource dependence. We have shown the framework has many uses, including estimating key ecological parameters, forecasting steady state populations and predicting environmentally relevant thresholds (in this case, tipping points). In our Antarctic shallow-water case study, we were able to use the model framework to demonstrate that all seven sites are either in the vicinity of, or have already passed, the tipping point, and are at risk of endemic invertebrate communities being replaced by beds of macroalgae, which is an ecologically significant finding. Beyond this case study, the mathematical framework we have introduced is highly flexible and could be applied to any biological or ecological system where there is a dependence of the system state on the available energy in the form of nutrients or some other fundamental requirement. 

\section*{Software and data availability}

The code used in this work was written in MATLAB (R2021b) and is available to download from the Figshare Repository \url{https://doi.org/10.6084/m9.figshare.25458229}. This code comprises 44 MATLAB code files which perform the analysis and produces the figures shown in this manuscript. Also included is a MATLAB data file containing results generated by the analysis (including the light time series datasets, estimated annual light doses, averaged algae data, model outputs and credible intervals). 

The code, data files and model output files contained in this repository are sufficient to generate all results shown in the paper (including all of the results shown in the Supplementary Material). Additional data fed into the Sequential Monte Carlo algorithm to generate the posterior distributions are confidential and therefore have not been provided in this repository, but are available upon reasonable request. 

\section*{Declaration of competing interest}

The authors declare that they have no known competing financial interests or personal relationships that could have appeared to influence the work in this paper. 

\section*{Acknowledgements}

The authors thank Patrick Hassard, Sarika Karanth, Jonathan Stark, and Sarah Vollert for helpful discussions during preparation of this manuscript. ELM and GFC were supported by the Australian Research Council (ARC) SRIEAS Grant SR200100005 Securing Antarctica’s Environmental Future. MPA's contribution was funded by an ARC Discovery Early Career Researcher Award DE200100683.

\clearpage 
\bibliographystyle{elsarticle-harv} 
\bibliography{AntarcticaBib}

\end{document}